\documentclass[twocolumn]{aa}  
\usepackage{natbib,twoopt}
\usepackage{hyperref}
\hypersetup{colorlinks=true,linkcolor=blue,citecolor=blue,filecolor=magenta,urlcolor=cyan,}
\bibpunct{(}{)}{;}{a}{}{,}             %% natbib format for A&A and ApJ
\usepackage{graphicx}
\usepackage{txfonts}
\def\aap{{\em A}\&{\em A}}

\def\apjl{{\em ApJ}}
\def\apjs{{\em ApJS}}

\newcommand{\kms}{\ifmmode {\rm km\ s}^{-1} \else km s$^{-1}$\ \fi}
\newcommand{\ergs}{\ifmmode {\rm erg\ s}^{-1} \else erg s$^{-1}$\ \fi}
\newcommand{\lb}{\ifmmode L_{\rm Bol} \else $L_{\rm Bol}$\ \fi}
\newcommand{\ledd}{\ifmmode L_{\rm Edd} \else $L_{\rm Edd}$\ \fi}
\newcommand{\lx}{\ifmmode L_{\rm 2-10keV} \else  $L_{\rm 2-10keV}$\ \fi}
\newcommand{\ha}{\hbox{H$\alpha$}}
\newcommand{\hb}{\hbox{H$\beta$}}

\newcommand{\mbh}{\ifmmode M_{\rm BH}  \else $M_{\rm BH}$\ \fi}
\newcommand{\lv}{\ifmmode \lambda L_{\lambda}(5100\Ang) \else $\lambda L_{\lambda}(5100\Ang)$\ \fi}
\newcommand{\lbol}{\ifmmode L_{\rm Bol} \else $L_{\rm Bol}$\ \fi}

\newcommand{\ci}{[\rm C {\sc I}]\ }

\newcommand{\oi}{\hbox{[O\,{\sc i}]}}

\newcommand{\nii}{\hbox{[N\,{\sc ii}]}}

\newcommand{\oiii}{\hbox{[O\,{\sc iii}]}}
\newcommand{\heii}{\hbox{He\,{\sc ii\ }}}

\newcommand{\msun}{ M_{\odot}}
\newcommand{\hi}{{\rm HI}}
\newcommand{\hii}{\hbox{H\,{\sc ii}}}

\newcommand{\oh}{\ifmmode 12+ \log({\rm O/H}) \else 12+log(O/H) \fi}
\newcommand{\mdot}{\ifmmode \dot{m} \else \dot{m} \fi }
\newcommand{\llog}{\ifmmode {\rm log} \else {\rm log} \fi }
\newcommand{\Ang}{\mathring{\mathrm{A}}}
\begin{document}

   \titlerunning{Molecular Gas in Haro 11}
   \authorrunning{Gao et al. 2022}

   \title{The molecular gas resolved by ALMA in the low-metallicity dwarf merging galaxy Haro 11}

   \author{Yulong Gao\inst{1,2}, Qiusheng Gu\inst{1,2}, Yong Shi\inst{1,2}, Luwenjia Zhou\inst{1,2}, Min Bao\inst{1,3,4}, Xiaoling Yu\inst{1,2}, Zhiyu Zhang\inst{1,2}, Tao Wang\inst{1,2}, Suzanne C. Madden\inst{5}, Matthew Hayes\inst{6}, Shiying Lu\inst{1,2}, and Ke Xu\inst{1,2}
          }

   \institute{School of Astronomy and Space Science, Nanjing University, Nanjing 210093, China \ \email{yulong@nju.edu.cn}
             \and
             Key Laboratory of Modern Astronomy and Astrophysics (Nanjing University), Ministry of Education, Nanjing 210093, China
             \and
             School of Physics and Technology, Nanjing Normal University, Nanjing 210023, China
             \and 
             Institute of Astronomy, The University of Tokyo, Osawa 2-21-1, Mitaka, Tokyo 181-0015, Japan
             \and
             CEA, Irfu, DAp, AIM, Universit\`e Paris-Saclay, Universit\`e de Paris, CNRS, F-91191 Gif-sur-Yvette, France
             \and 
             Stockholm University, Department of Astronomy and Oskar Klein Centre for Cosmoparticle Physics, AlbaNova University Centre, SE-10691, Stockholm, Sweden
             }

   %\date{Received September 15, 1996; accepted March 16, 1997}

% \abstract{}{}{}{}{} 
% 5 {} token are mandatory
 
  \abstract
   {The physical mechanisms for starburst or quenching in less massive ($M_* \leq 10^{10} \msun$) galaxies are unclear. The merger is one of the inescapable processes referred to as both starburst and quenching in massive galaxies. However, the effects of the merger on star formation in dwarf galaxies and their evolution results are still uncertain.}
   {We aim to explore how to trigger and quench star formation in dwarf galaxies by studying the metal-poor gas-rich dwarf mergers based on the multi-band observations at a spatial resolution of $\sim$ 460 pc. }
   {We use the archival data of ALMA (band 3, 8) and VLT/MUSE to map CO($J=$1-0), [CI]($^3$P$_1 - ^3$P$_0$), and $\ha$ emission in one of the most extreme starburst merging dwarf galaxies, Haro 11.}
   {We find the molecular gas is assembled around the central two star-forming regions (knots B and C).
   The molecular/ionized gas and stellar components show complex kinematics, indicating that the gas is probably at a combined stage of collision of clouds and feedback from star formation. The peak location and distribution of [CI](1-0) strongly resemble the CO(1-0) emission, meaning that it might trace the same molecular gas as CO in such a dwarf merger starburst galaxy. The enhancement of line ratios ($\sim 0.5$) of [CI]/CO around knot C is probably generated by the dissociation of CO molecules by cosmic rays and far-ultraviolet photons. Globally, Haro 11 and its star-forming regions share similar SFEs as the high-$z$ starburst galaxies or the clumps in nearby (U)LIRGs.}
   {Given the high SFE, sSFR, small stellar mass, low metallicity, and deficient $\hi$ gas, Haro 11 could be an analog of high-$z$ dwarf starburst and the potential progenitor of the nearby less massive elliptical galaxies. The significantly smaller turbulent pressure and viral parameter will probably trigger the intense starbursts. We also predict that it will quench at $M_* \leq 8.5 \times 10^9 \ \msun$.}

   \maketitle
%
%-------------------------------------------------------------------
\section{Introduction}

The physical causes of the starburst and quenching in galaxies are open questions. In the star formation rate (SFR) versus stellar mass ($M_*$) space, galaxies distribute as two broad categories. Star-forming galaxies (SFGs, also known as the blue clouds), most of which are spiral or late-type galaxies, lie on an almost linear relation \citep[main sequence, e.g., ][]{Brinchmann2004a, Daddi2007, Peng2010, Lilly2013, Speagle2014}. Starburst galaxies undergo an exceptionally intense phase of star formation, which are above the main sequence, representing an evolutionary phase in the life of a galaxy \citep[e.g., ][]{elbaz2018, Orlitova2020}. Quiescent galaxies (QGs, red sequences), most elliptical or early-type galaxies, have significantly lower SFRs than SFGs and thus lie below the main sequence.  Some so-called ``green valley'' galaxies lie between the blue SFGs and red QGs, regraded as the transition zone or quenching galaxies \citep[e.g., ][]{Bell2003, Fang2012}.  These galaxies above/below the main sequence imply that SFGs might enhance or shut down their star formation at some point during their lifetime.

A variety of processes can induce starburst activities in galaxies. The strong tidal force caused by gravitational interactions between close pairs or merger galaxies will perturb the orbits of gas and stars, triggering the gas flowing toward the galaxy center efficiently \citep[e.g., ][]{Larson1978Jan, DiMatteo2007Jun, Ellison2008}. In spiral-barred galaxies, because of the gravitational instabilities of the stellar bar, the gas in the outer spiral can also be transferred into the central region and form new stars \citep[e.g.,][]{Hopkins2009, Elmegreen2009a, Emsellem2015, DiazGarcia2021}. Similarly, the star formation quenching is correlated with active galactic nuclei (AGN) feedback \citep[e.g.,][]{dimatteo2005a,kauffmann2007,fabian2012,Cheung2016May,kaviraj2017}, secular processes, such as stellar bars, oval disks and spiral structures \citep[e.g.,][]{kormendy2004,masters2011,cheung2013}. Meanwhile, some external processes, for example, the galaxy environment and the major mergers \citep[e.g.,][]{dimatteo2005a,hopkins2006,hopkins2008,knobel2015,peng2015,Weigel2017} can also affect the gas consumption and lead to the quenching of star formation.

The merging of galaxies is one of the inescapable processes in galaxy evolution, for both the starburst and quenching activities \citep{Hopkins2009}. By perturbing gas rotation, mergers will lead to the rapid inflows of gas and power the intense starburst or feed the central massive black hole. The energetic feedback from AGN or starbursts will heat the interstellar medium (ISM), preventing the gas from cooling and sweep out gas from their host galaxies \citep[e.g., ][]{fabian2012, Cheung2016May,harrison2018}, which will suppress the star formation and convert the blue SFGs into quiescent red massive elliptical galaxies. In the local universe, major mergers are always observed between spiral galaxies, which could result in the ultraluminous infrared galaxies \citep[(U)LIRGs, e.g.,][]{Papadopoulos2007, Israel2015, Espada2018, Spence2018, Shangguan2019a}. At high redshift universe, where galaxy mergers were more prevalent, most of the observed merger cases are bright and massive submillimeter galaxies (SMGs) \citep[e.g.,][]{alexander2005,shapiro2008,narayanan2010,brisbin2017,wardlow2017}, suggested to be the progenitors of massive quiescent elliptical galaxies in the local universe. 

However, the mechanisms of triggering or quenching star formation in low stellar mass or dwarf galaxies \citep[i.e, $M_* < 5 \times 10^9 \ \msun$, ][]{Stierwalt2015} are unclear. WHile a large fraction ($> 70\%$) in the local universe are dwarf galaxies, only a few percent of them are starburst galaxies, including blue compact galaxies (BCGs), Lyman-$\alpha$ reference sample \citep[LARS, ][]{hayes2013,ostlin2014}, Lyman-break analogs \citep[LBAs, ][]{heckman2001,heckman2005}, and Green Peas  \citep[see the review in][]{Orlitova2020}. Many BCGs are relatively isolated, while the morphologies and velocity fields implicate a recent interaction with neighbors. Furthermore, the stellar components in BCGs contain a mixture of young and old populations \citep{kunth2000}, perhaps indicating that interactions are responsible for the starburst activity in BCGs. However, because of the weak gravitational potential and strong turbulence by stellar winds, the conversion from molecular gas to stars is suggested to be inefficient in previous star formation models. Whether the merging of two dwarf galaxies can trigger starburst activity, is still an open question. Recently, \cite{Zhang2020h, Zhang2020g} reported that the star cluster formation rate ($\propto$ SFR) in the galaxy VCC 848, a remnant of a gas-rich dwarf-dwarf merger, is enhanced $\sim$ 1.0 dex during the past $\sim$ 1 Gyr relative to its earlier times.

Though the dwarf-dwarf mergers are quite rare \citep[$\sim 3\%$, ][]{KadoFong2020} within dwarf galaxies in local universe, they are expected to be frequent at high redshift \citep[e.g.,][]{Fensch2017,Husko2021,romano2021,Whitney2021}. The evolution results of these dwarf mergers are still uncertain. \cite{geha2012} performed a systematic study about the quenched field galaxy fraction using the SDSS DR8 spectra, demonstrating that it is smaller than 0.06$\%$ when $M_* < 10^9 \ \msun$, while are about 0.5$\%$ and 3$\%$ for less massive galaxies at $M_* < 10^{9.5} \ \msun$ and $M_* < 10^{9.9} \ \msun$, respectively.  Assuming the merger quenching scenario in dwarf galaxies is similar to more massive ones, dwarf mergers at high redshift would be the progenitors of these nearby dwarf elliptical galaxies. However, it is difficult to study the detailed starburst activities and gas consumption for these distant dwarf mergers because of the low brightness. Thus, analogs of these distant cases, metal-poor gas-rich dwarf mergers with high spatial resolution, multi-band observations, can provide the ideal laboratories to study the physical mechanisms of triggering and quenching the star formation in such dwarf galaxies.

In this paper, we focus on the molecular/neutral gas properties of Haro 11, which is a merging system and one of the most extreme starburst BCDs in the nearby universe \citep[e.g.,][]{Cormier2012,Cormier2014,Oestlin2015,Pardy2016,Menacho2019,Menacho2021,Oestlin2021}. The redshift ($z$) is about 0.0206, collected from the NASA/IPAC Extragalactic Database\footnote{\url{http://ned.ipac.caltech.edu}} (NED), corresponding to the distance of 87.2 Mpc and a scale of 420 pc per arc second.
Haro 11 is a late-stage merger, showing similar morphology and kinematics of stars and ionized gas to the well-known Antennae (NGC 4038/4039) galaxy \citep{Oestlin2015}. The starburst activities mainly occur at three bright knots: A (south-western), B (central), and C (eastern), shown in Fig. \ref{fig:Ha_CO10}. These knots are found to contain more than 200 massive star clusters \citep[e.g.,][]{adamo2010a,Oestlin2015,Menacho2021}. Stellar feedback produced by intense starbursts has created outflows and multi-scale structures, such as filaments, arcs and bubbles, which make the merger kinematics much more complex \citep{Menacho2021}. The gas-phase metallicities ($\oh$, $Z$), calculated from the electron temperature method, are about 8.09 (0.24 $Z_{\odot}$), 8.25 (0.35 $Z_{\odot}$) and 7.80 (0.12 $Z_{\odot}$) for knots A, B and C, respectively \citep{james2013}. The total SFR is about $25.1^{+28.6}_{-13.4} \ \msun \ \rm yr^{-1}$ based on the $\ha$ and infrared luminosity \citep{remy-ruyer2015}. The molecular gas tracers CO(1-0), CO(2-1), and CO(3-2), detected by ATNF Mopra 22-m and APEX telescopes determined that the total H$_2$ gas mass ranges from $2.5 \times 10^8 \ \msun$ using a Galactic conversion factor, $X_{\rm CO}$ \citep{Cormier2014}.
The dust mass derived from the dust Spectral Energy Distribution modeling of the mid-infrared to submillimeter observations which included \textit{Spitzer} and \textit{Herschel} data, is $9.9 \times 10^6 \ \msun$ \citep{remy-ruyer2015}.
\cite{Pardy2016} presented the first robust detection of $\hi$ 21 cm emission for Haro 11 using the GBT telescope, and found the HI mass to be about $5.1 \times 10^8 \ \msun$. \cite{Cormier2014}, and \cite{Pardy2016} reported that the gas in Haro 11 is dominated by molecular gas instead of $\hi$ gas. \cite{Cormier2014} also found the gas depletion of Haro 11 is very rapid, about 0.1 Gyr, indicating that it is efficient in forming stars. However, the presence of additional molecular gas not traced by CO (CO-dark molecular gas) proposed by \cite{madden2020}, reduces the SFR efficiency.

However, limited by the large beam size of the single antenna in previous studies, we cannot answer the following questions:
\begin{itemize}
    \item What is the distribution of molecular gas in Haro 11?
    \item What conditions within these individual gas clouds encourage the presence of starbursts?
    \item How does the star formation efficiency vary within the galaxy? 
    \item How does the gas depletion compare to the star-forming regions in massive merger systems, e.g., (U)LIRGs?
    \item What is the role of different gas phases in such a dwarf merger system?
    \item What is the impact of the starburst on the surrounding molecular and atomic gas? 
    \item How does the conversion from HI to H$_2$ in the dwarf mergers compare to massive mergers and other isolated galaxies?
\end{itemize}
To address these questions We collect and analyse the high-resolution sub/millimeter ALMA observation of CO($J=$1-0) (hereafter CO(1-0)) and [CI]($^3$P$_1 - ^3$P$_0$) (hereafter [CI](1-0)) emission lines and the archival optical VLT/MUSE data and investigate the starburst activities and gas consumption in Haro 11 at 500 pc scales.

The organisation of the paper is the following: In Section 2, we present the observations and data reduction, then derive the properties of the molecular gas and star formation activities. The main results and comparison with other galaxies are presented in Section 3 and 4, respectively, along with a summary in Section 5. Throughout this paper, we assume a flat $\Lambda$CDM cosmology model, $\Omega_\Lambda=0.7$, $\Omega_{\rm m}=0.3$, and $H_0=70$ km s$^{-1}$ Mpc$^{-1}$. We adopt the solar metallicity ($Z_{\odot}$) as $\oh = 8.69$ \citep{Allende2001}.

%--------------------------------------------------------------------
\section{Observations and Data Reduction}
\label{sec:data}

\subsection{ALMA data}
\label{subsec:alma_data}

This paper collects the band 3 and band 8 observations (ID: 2013.1.00350.S, PI: Cormier; 2017.1.01457.S, PI: Hayes) from the ALMA data archive. The band 3 observations (12-m array) were obtained on 7 August 2016, with a total integration time of 2903 seconds and a field of view (FoV) of $\sim 55\arcsec$. The frequency ranges from 112.0 to 113.8 GHz, covering the CO(1-0) emission line. The longer integration time (8981 seconds) observations in band 8 (12-m array) were performed on 13 May 2018, covering an FoV of $\sim 12\arcsec$. The frequency (481 -- 483 GHz) covers the $\ci$(1-0) emission line. The largest angular scales are about 4.2$\arcsec$ and 5.3$\arcsec$ for band 3 and band 8 observations, respectively. We use the Common Astronomy Software Applications (CASA) package \citep[version 5.7, ][]{McMullin2007a} to perform the standard calibration. The ALMA data were cleaned with the natural weighting with the CASA task \textit{tclean} and were primary-beam corrected. The synthesized beam sizes in the CO(1-0) and $\ci$(1-0) final data cubes are about $1.1\arcsec \times 1.0\arcsec$ (462 $\times$ 420 pc) and $0.5\arcsec \times 0.5\arcsec$ (210 $\times$ 210 pc), respectively. The pixel sizes in CO and $\ci$ images are about 0.1$\arcsec$. The RMS values are about 1.2 mJy/beam and 1.3 mJy/beam for CO(1-0) and $\ci$(1-0) data at a channel width of 5 \kms, respectively. We create the CO and $\ci$ moment maps with the CASA task \textit{immoments} by applying a threshold of 4 $\times$ the RMS.

%-------------------------------------------------------------
%                                             Simple A&A Table
%-------------------------------------------------------------
%
\begin{table}[t]
\scriptsize
\caption{Properties of Haro 11}             % title of Table
\label{table:1}      % is used to refer this table in the text
\centering                          % used for centering table
\begin{tabular}{l l l c}        % centered columns (4 columns)
\hline\hline                 % inserts double horizontal lines
Property & Unit & Value & Reference \\    % table heading 
\hline                        % inserts single horizontal line
Other name & ... & ESO 350-IG38 & 1 \\      % inserting body of the table
RA (J2000) & h:m:s &  00:36:52.7 & 1 \\
Dec (J2000) & d:m:s & -33:33:17 & 1 \\
$V_{\rm sys}$(optical) & \kms & 6194 $\pm$ 5.1 & 2 \\
Redshift & ... & 0.0206467 & 2 \\
Distance & Mpc & 87.2 $\pm$ 6.1 & 3 \\
Scale & pc/arcsec & $\sim$ 420 & 3 \\
$M_*$ & $10^9 \ \msun$ & 2.6 $\pm$ 0.8 & 4 \\
$V_{\rm sys}$ (CO) & \kms & 6051.3 $\pm$ 2.1 & 4 \\
CO(1-0) flux & Jy \kms & 3.3 $\pm$ 0.1 & 4 \\
$\ci$ flux & Jy \kms & 7.4 $\pm$ 0.3 & 4 \\
$M_{\rm H_2}$(CO) & $10^9 \ \msun$ & 3.8$^{+3.3}_{-3.1}$ & 4 \\
$M_{\rm H_2}$([CI]) & $10^8 \ \msun$ & 2.1-5.3 & 4 \\
$M_{\hi}$ & $ 10^8 \ \msun$ & 5.1 & 5 \\
$\rm SFR_{\ha+\rm TIR}$ & $\msun \ \rm yr^{-1}$ & $25.1^{+28.6}_{-13.4}$ & 6 \\
$\rm SFR_{{\ha}(HST)}$ & $\msun \ \rm yr^{-1}$ & 24.0 & 7 \\
$\rm SFR_{\rm TIR}$ & $\msun \ \rm yr^{-1}$ & 28.6  & 8 \\
$\rm SFR_{\ha}$ & $\msun \ \rm yr^{-1}$ & 25.1 $\pm$ 0.4 & 4 \\
Size & kpc$^2$ & 12.5 & 4 \\
$\sigma_{v}$ & \kms & 17.2$^{+8.3}_{-11.5}$ & 4 \\
$P_{\rm turb}$ & $10^6 \rm \ K \ cm^{-3}$ & 3.4$^{+10.2}_{-3.2}$ & 4 \\
$\alpha_{\rm vir}$ & ... & 0.17$^{+0.11}_{-0.06}$ & 4 \\
\hline                                   
\end{tabular}
\tablefoot{The \cite{Kroupa2001} IMF is used to derive the stellar mass and SFR in this work. We assume the CO-H$_2$ conversion factor $\alpha_{\rm CO}$ as $60 \ \msun \rm (K \ \kms \ {pc}^{-2})^{-1}$, with a range of $10 - 110 \ \msun \rm (K \ \kms \ {pc}^{-2})^{-1}$. We use the python packages \textit{Statmorph} to estimate the radius $R_{\rm 80}$ containing 80$\%$ flux in $r$ band, then calculate the size of Haro 11. The CO(1-0) velocity dispersion $\sigma_{v}$, internal turbulent pressure $P_{\rm turb}$ and the viral parameter $\alpha_{\rm vir}$ are shown as the median values with one sigma ranges. We also use the $\ci$ luminosity to derive the molecular gas based on the calibration in \cite{madden2020} at metallicity of 0.25 $Z_{\odot}$ and 0.1 $Z_{\odot}$, respectively, with a systemic uncertainty of 0.3 dex.}
\tablebib{
(1) \cite{Oestlin2015}; (2) \cite{Menacho2021}; (3) NED; (4) This work; (5) \cite{Pardy2016}; (6) \cite{remy-ruyer2015}; (7) \cite{hayes2007}; (8) \cite{madden2013}.
}
\end{table}

\begin{table*}[t]
\scriptsize
\caption{Properties of star formation knots}             % title of Table
\label{table:2}      % is used to refer this table in the text
\centering                          % used for centering table
\begin{tabular}{l r r c c c c c c c c}        % centered columns
\hline\hline                 % inserts double horizontal lines
Property & RA (J2000) & Dec (J2000) & Diameter & $\oh$ & $M_*$ & SFR & $M_{\rm H_2}$ & $\sigma_{v}$ & $P_{\rm turb}$ & $\alpha_{\rm vir}$ \\
... & deg & deg & arcsec (pc) & ... & $10^7\ \msun$ & $\msun \rm \ yr^{-1}$ & $10^8\ \msun$ & \kms & $\rm 10^6 \ K \ cm^{-3}$ & ... \\
    % table heading 
\hline                        % inserts single horizontal line
A & 9.2177732 & -33.555504 & 1.7 (710.7) & 8.09 $\pm$ 0.20 & 4.3 $\pm$ 1.2 & 1.42 $\pm$ 0.03 & $< 2.3$ & ... & ... & ... \\
B & 9.2182527 & -33.554668 & 3.0 (1254.2) & 8.25 $\pm$ 0.15 & 8.9 $\pm$ 2.6 & 10.92 $\pm$ 0.08 & 23$^{+20}_{-19}$ & 23.8$^{+3.7}_{-12.1}$ & 8.7$^{+16.9}_{-8.1}$ & 0.17$^{+0.17}_{-0.06}$ \\
C & 9.2192717 & -33.554783 & 1.1 (459.8) & 7.80 $\pm$ 0.13 & 19.1 $\pm$ 5.5 & 1.09 $\pm$ 0.02 & 1.1$^{+1.0}_{-0.9}$ & 16.2$^{+6.6}_{-10.8}$ & 1.7$^{+4.0}_{-1.1}$ & 0.15$^{+0.09}_{-0.11}$ \\
T & 9.2174637 & -33.554432 & 0.9 (376.3) & ... & 0.3 $\pm$ 0.1 & 0.31 $\pm$ 0.02 & $< 0.1$ & ... & ... & ... \\
\hline                                   
\end{tabular}
\tablefoot{The metallicities ($\oh$) are determined with electron temperature method using the $\oiii\lambda4363$ emission line, collected from \cite{james2013}. The CO(1-0) velocity dispersion $\sigma_{v}$, internal turbulent pressure $P_{\rm turb}$ and the viral parameter $\alpha_{\rm vir}$ are shown as median values and one sigma ranges, respectively.}
\end{table*}

\begin{figure*}[t]
   \centering
   \includegraphics[width=0.45\textwidth]{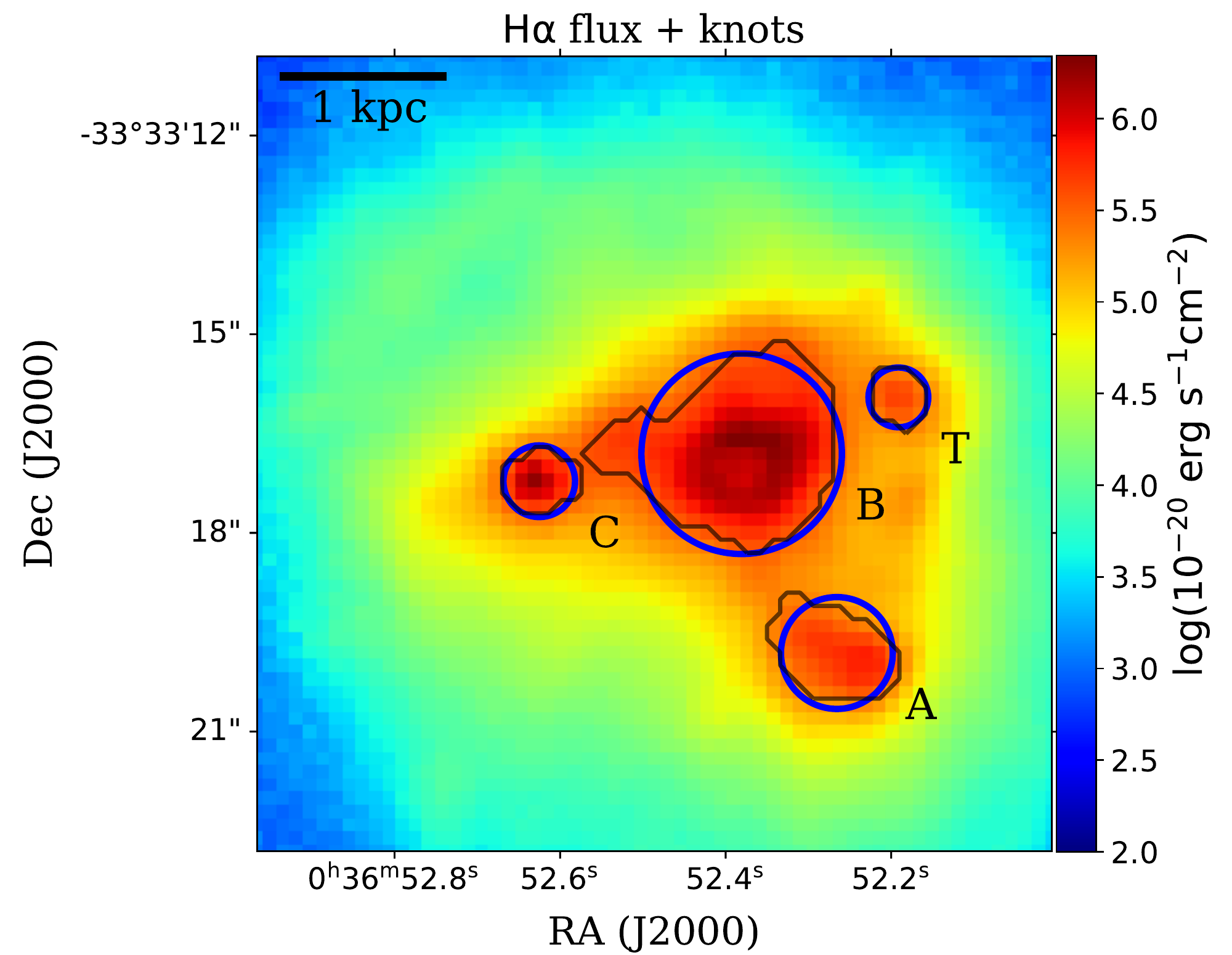}
   \includegraphics[width=0.44\textwidth]{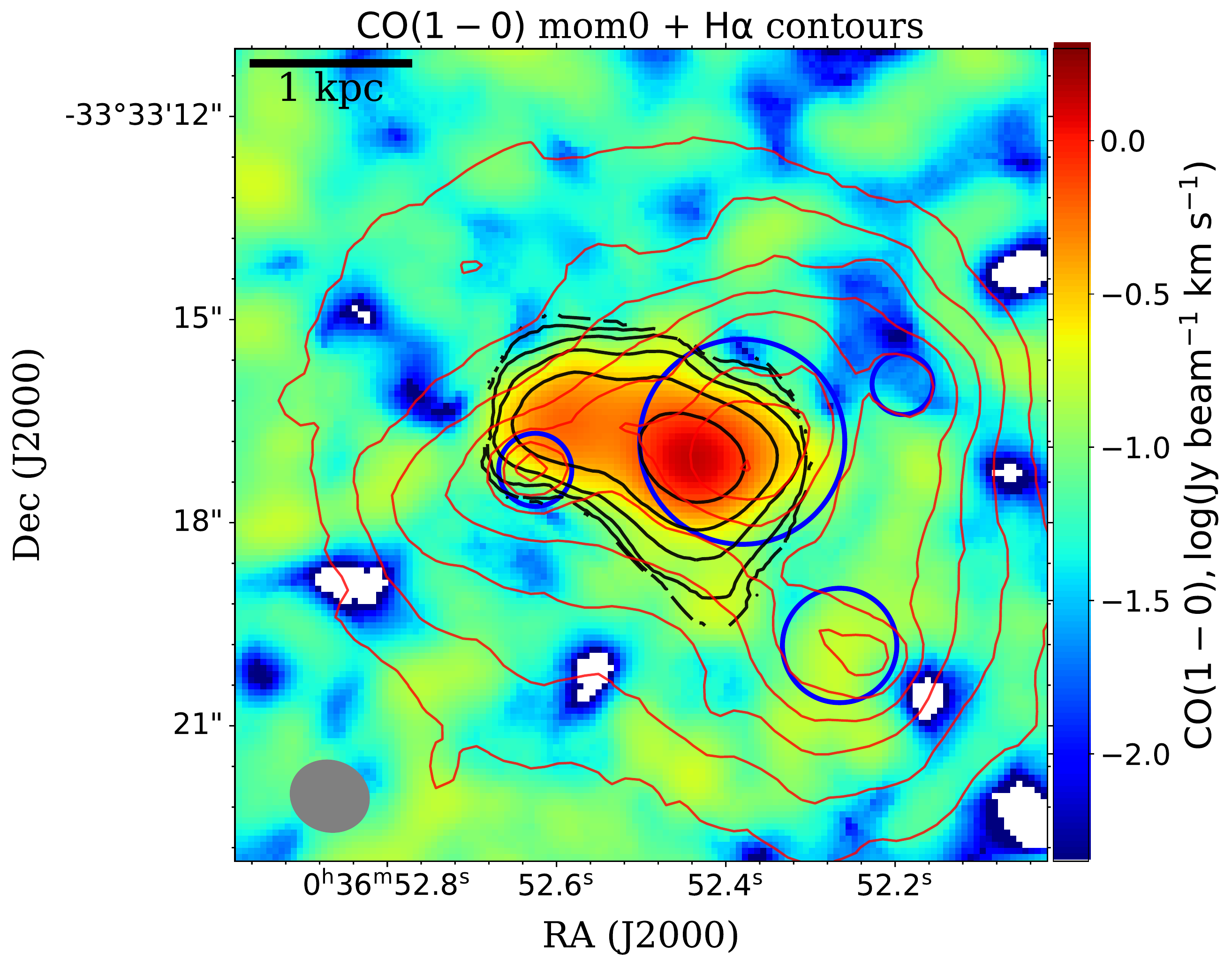}
   \caption{$Left$: the integrated intensity map of attenuation-corrected $\ha$ emission. The gray contours represent the identified star formation knots by \textit{Astrodendro}, and the blue circles mean the same area size. $Right$: the integrated intensity (moment 0) of CO(1-0) emission line without threshold setting. The black contours represent the integrated intensity of CO(1-0) with a 4$\times$RMS threshold. The $\ha$ intensity is marked by red contours, with levels of $(1/2)^n$ ($n \ =$ 1, 2, ..., 8) times the maximum value of $\ha$ intensity. The CO(1-0) beam size is shown as a solid gray elliptical circle in the bottom left.}
   \label{fig:Ha_CO10}
 \end{figure*}

\subsection{VLT/MUSE data}
\label{subsec:muse_data}

Haro 11 was observed by the VLT/MUSE \citep{Bacon2010} on December 2014 and August 2016 \citep{Menacho2019,Menacho2021}. The integration exposure time at the central $30\arcsec \times 30\arcsec$ region is about 3.1 hours. We collect the fully reduced data cube on the ESO archive website\footnote{\url{http://archive.eso.org/scienceportal/home}}. The seeing values during the observations are 0.6 -- 0.9$\arcsec$, the FWHM of the final image is about 0.8$\arcsec$ \citep{Menacho2019}. The rest-frame spectral range is $\rm 4750 - 9160 \Ang$ with a channel width of 1.25$\rm \Ang$. 

To derive the flux of pure emission lines, we employ the STARLIGHT package \citep{CidFernandes2005} to reproduce the stellar continuum. In this process, we assume the \cite{Chabrier2003} initial mass function (IMF), and perform a combination of 45 single stellar populations (SSPs) from \cite{Bruzual2003} model, which consist of three different metallicities and 15 stellar ages. The stellar mass within each spatial pixel can be derived from the SSP fitting results, with an uncertainty of 0.11 dex \citep{Bruzual2003, CidFernandes2005}. The total stellar mass is estimated by integrating the values in each pixel with continuum S/N larger than 5.0, and then converted to the \cite{Kroupa2001} IMF by adding 0.15 dex \citep{Bell2003,Shangguan2019a}. After subtracting the stellar continuum synthesis, we apply multiple Gaussians to fit these strong emission lines, such as $\ha$, $\hb$, and $\oiii\lambda\lambda4959,5007$. 

\subsection{Star formation knots}
\label{subsec:knots}

We can resolve Haro 11 into smaller star formation knots with high spatial resolution images from ALMA and MUSE. We use the python package {\textit{Astrodendro}\footnote{\url{https://dendrograms.readthedocs.io/en/stable/}}} \citep{Goodman2009Jan} to search for star-forming clumps in the $\ha$ map. This clump-finding algorithm is based on the dendrogram and has been used to identify reliable star-forming cores in galaxies (see \cite{Li2020c} for the detailed comparison of different clump-finding packages). To define the boundaries of clump structures, we need to specify a few parameters: $min\_value$, $min\_delta$ and $min\_npix$. The $min\_value$ constrains the minimum value in the field to be considered, $min\_delta$ represents the minimum significance of the structure to avoid including local maxima, and $min\_npix$ specifies the minimum number of pixels that a structure should contain. We only assume the $min\_npix$ as 16. The identified knots are shown in Fig. \ref{fig:Ha_CO10}, in which A, B and C are same as previous studies \cite[e.g.,][]{adamo2010a,Oestlin2015,Menacho2021}. Furthermore, the smaller star formation knot T (the name is same as the Table 3 of \cite{vader1993}), containing young massive star clusters, is also marked in Fig. \ref{fig:Ha_CO10} and is referred to as the ear between knots A and B \citep[e.g.,][]{Oestlin2015}.  

   \begin{figure*}[t]
   \centering
   \includegraphics[width=0.45\textwidth]{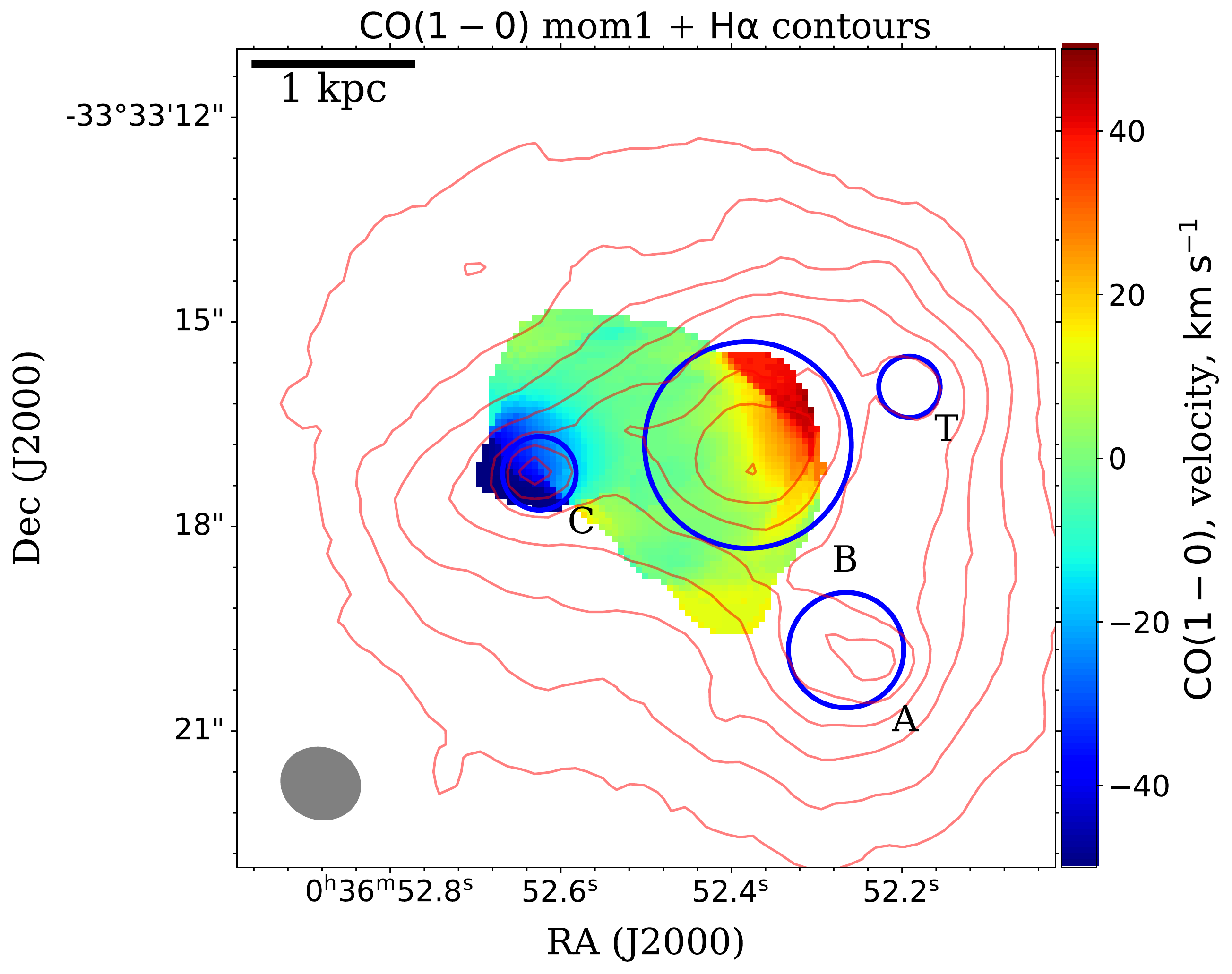}
   \includegraphics[width=0.45\textwidth]{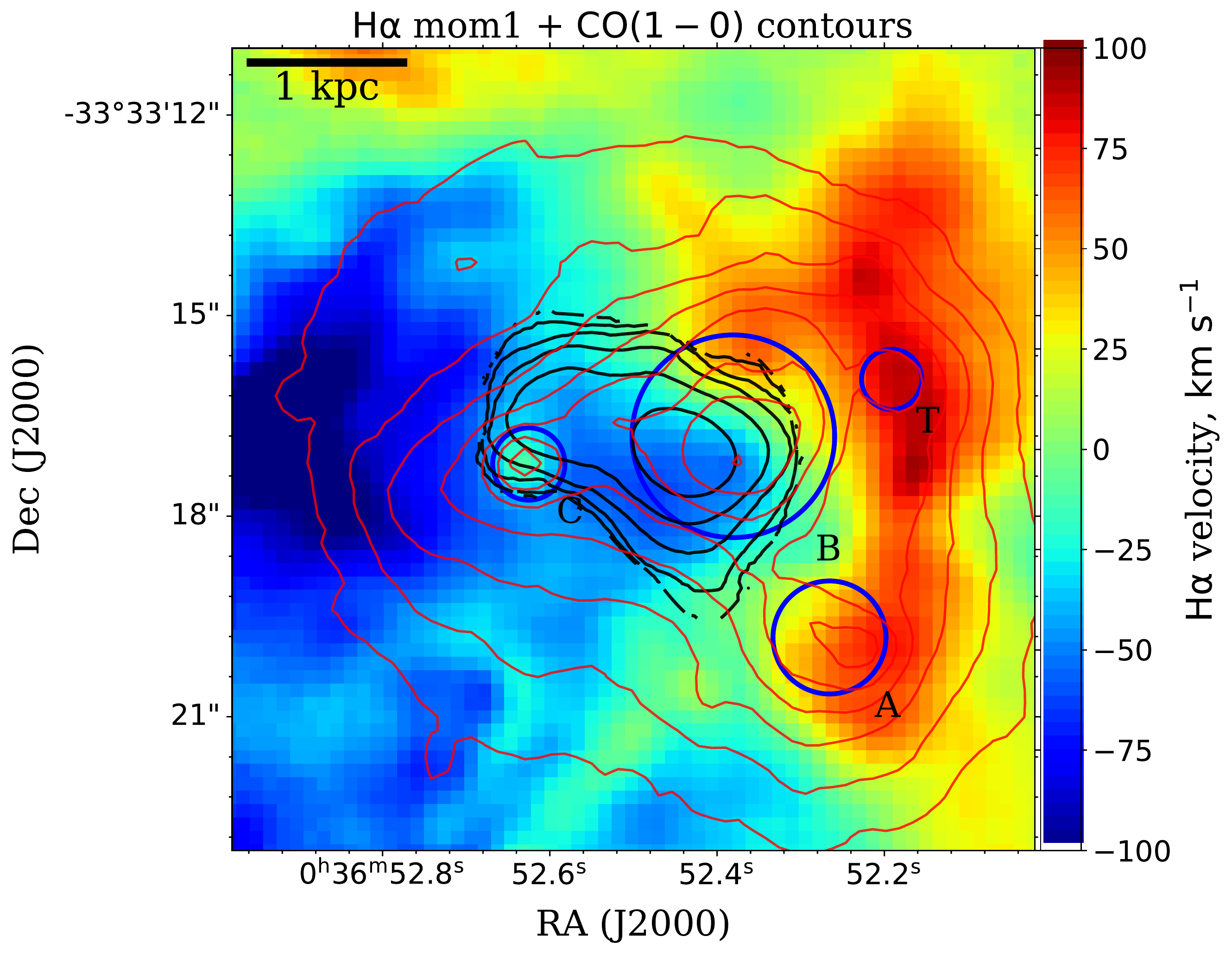}
   \includegraphics[width=0.45\textwidth]{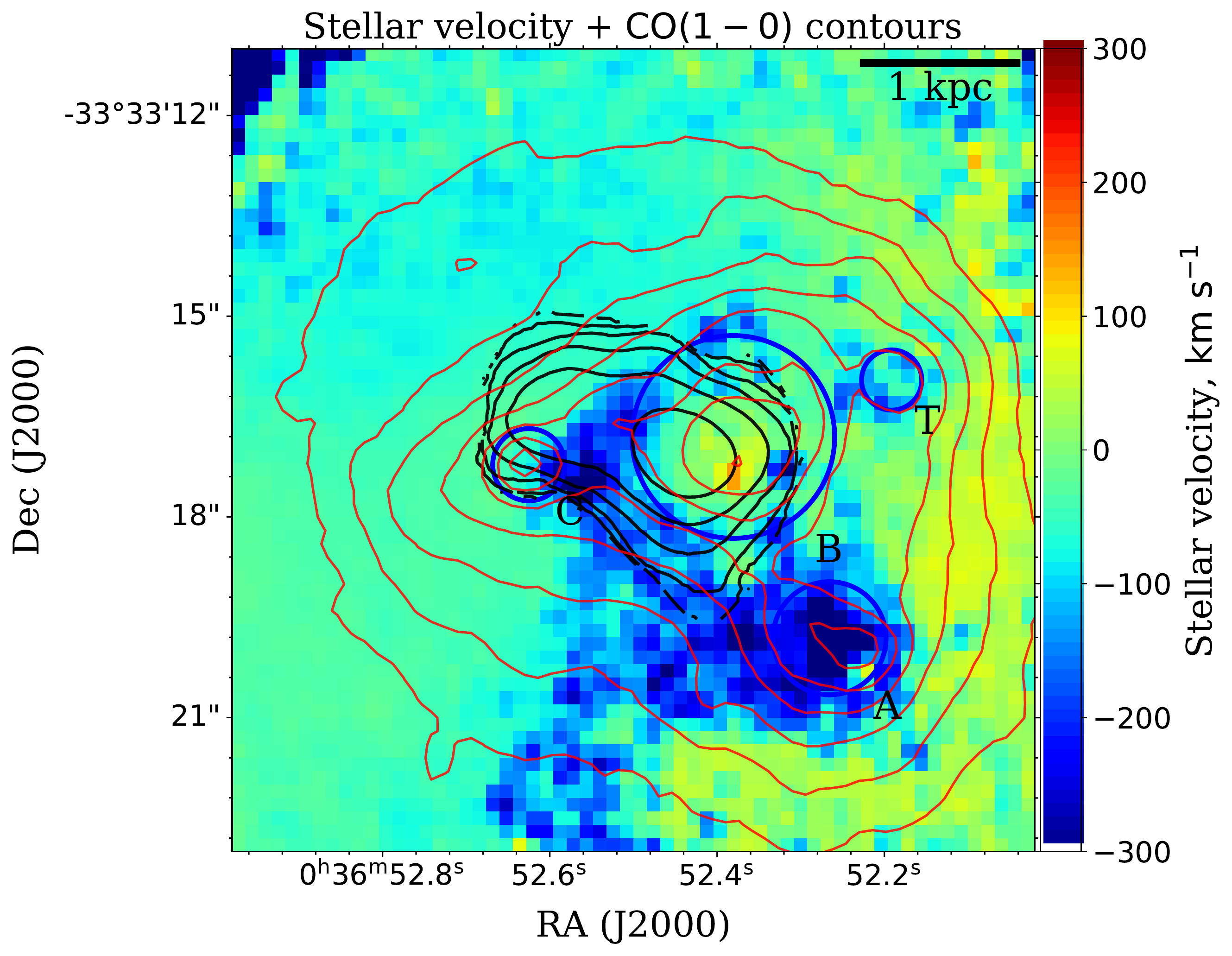}
   \caption{$Left$: the systemic velocity subtracted intensity-weighted velocity field (moment 1, 4$\times$RMS) of CO(1-0) emission line in Haro 11. $Right$: the systemic velocity subtracted intensity-weighted velocity field of $\ha$ emission. $Bottom$: the stellar velocity map derived from the SSP fitting results. Black contours and red contours represent the intensities of CO(1-0) and $\ha$ emission lines, respectively.}
   \label{fig:CO10_Ha_vel}
   \end{figure*}

   \begin{figure*}[t]
   \centering
   \includegraphics[width=0.45\textwidth]{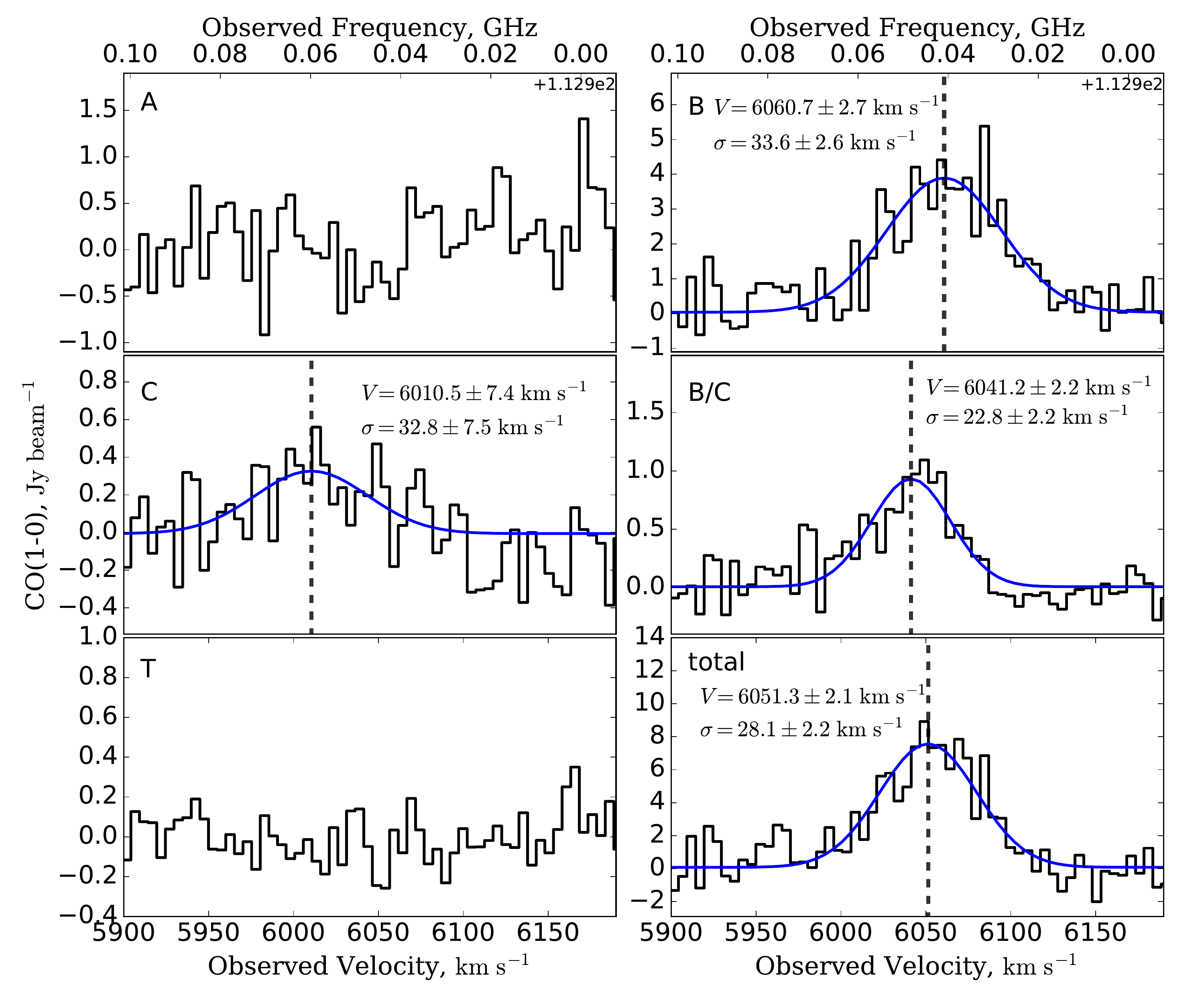}
   \includegraphics[width=0.45\textwidth]{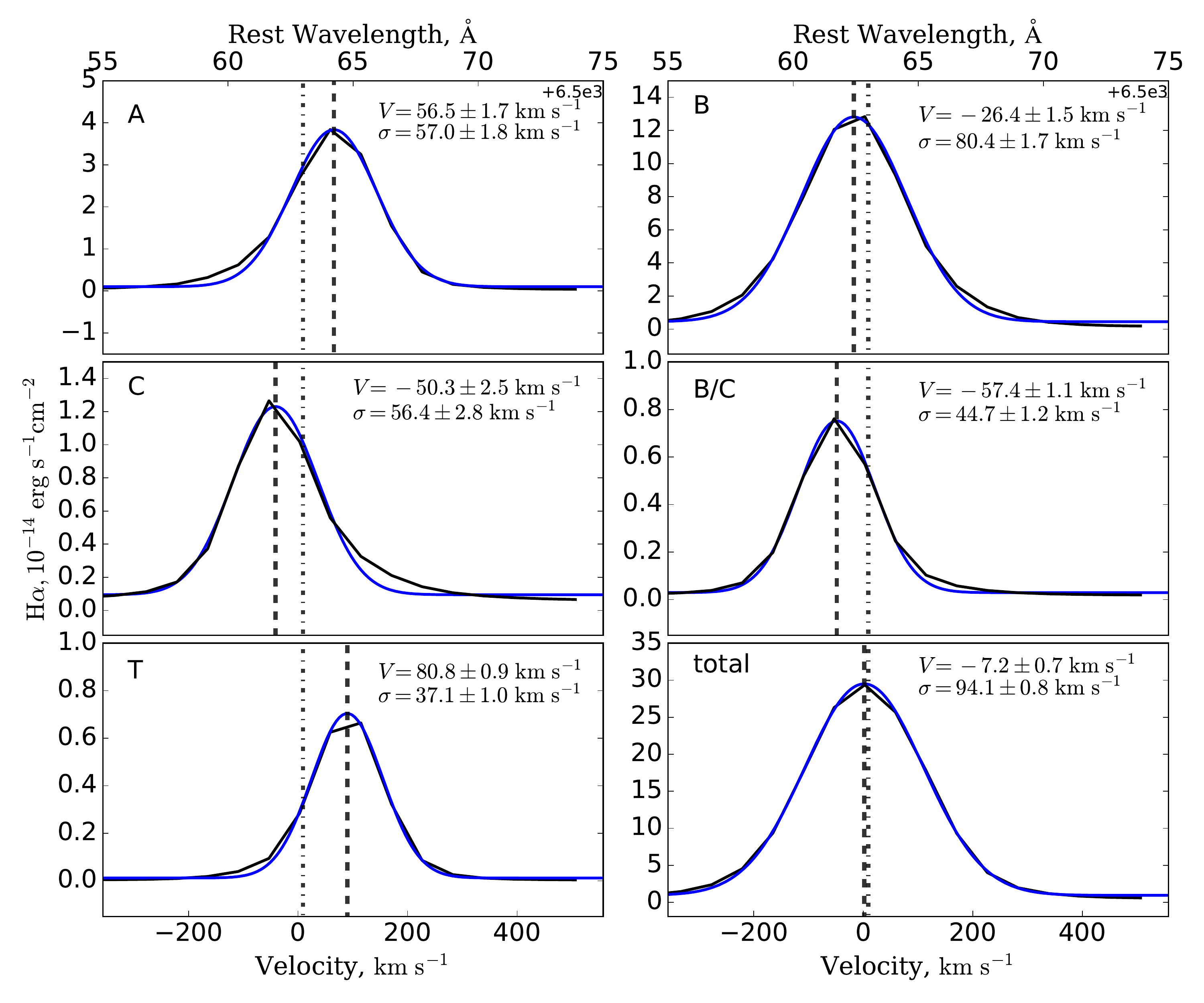}
   \caption{$Left$: the CO(1-0) spectra collected from the knots (A, B, C, T), region ($r \sim 0.6\arcsec$, marked as B/C) between B and C, as well as the entire CO region. $Right$: the $\ha$ spectra collected from the knots (A, B, C, T), region B/C, and the $R_{\rm 80}$ region. The dash-dotted lines show the rest-frame wavelength of the $\ha$ emission line. In each panel, blue lines represent the best one-Gaussian fitting spectra, attached with the relevant results, e.g., the velocity of CO and systemic velocity subtracted velocity of $\ha$. The velocity dispersion $\sigma$ values are derived from the Gaussian widths by subtracting the instrument resolution. Dashed lines represent the fitting velocity of CO and $\ha$.}
   \label{fig:CO10_Ha_spec}
   \end{figure*}

   \begin{figure*}[t]
   \centering
   \includegraphics[width=0.9\textwidth]{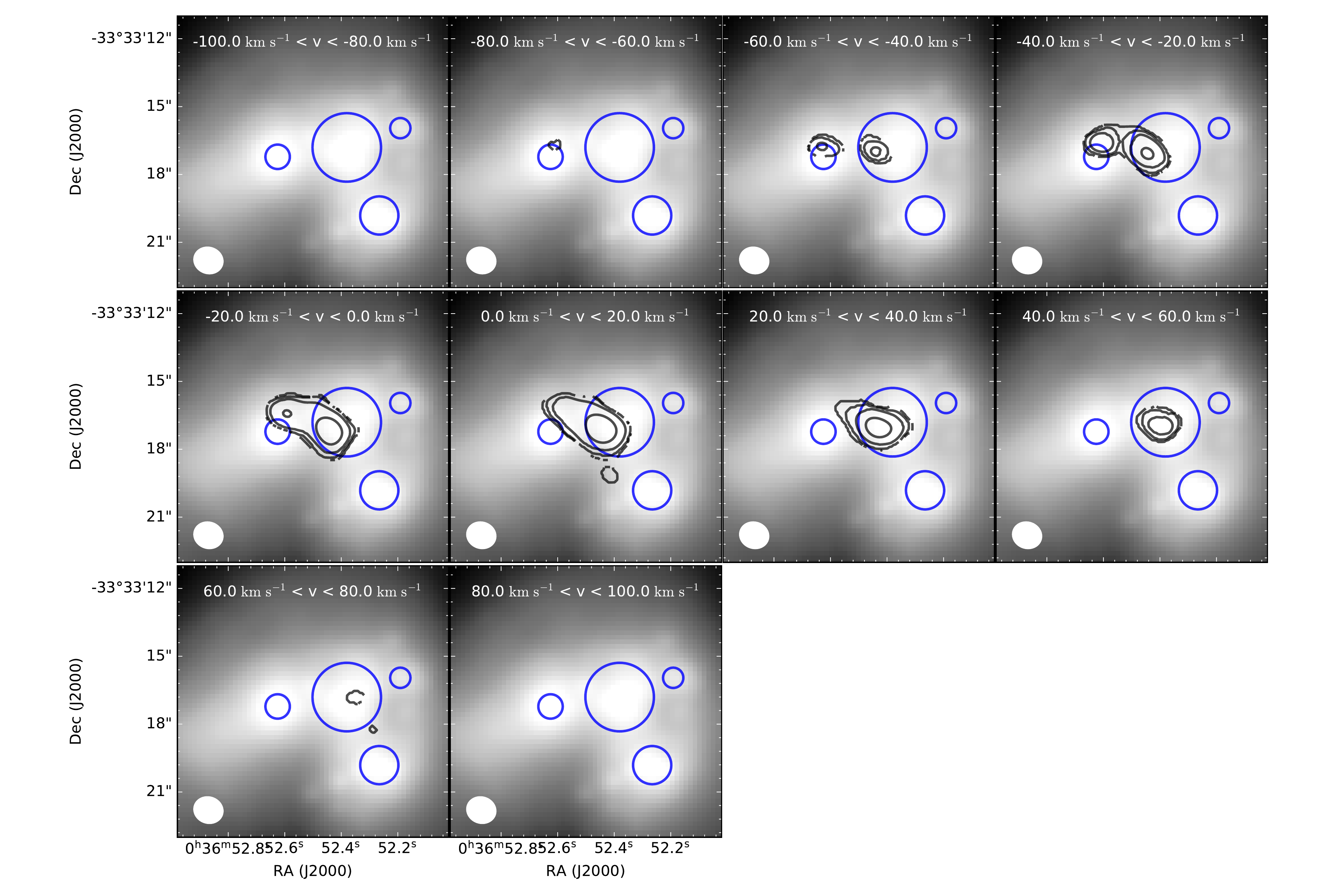}
   \caption{CO(1-0) channel maps of Haro 11 contoured on the optical stellar continuum image. The spectral channel width of the original CO(1-0) cube is rebinned to 20 $\kms$. The blue circles represent the four starburst knots, identical to Fig. \ref{fig:Ha_CO10}. The black contours show the intensity map at levels of 3, 6, 12 ... times of the RMS. The beam size of CO(1-0) is marked as the elliptical white circle in the bottom-left corner of each panel. }
   \label{fig:CO10_mom1_channels}
   \end{figure*}

\section{Results}
\label{sec:resu}

The spatial distribution of ionized and molecular gas can provide essential information about the gas kinematics and star formation activity. Here we describe the spatial distribution of the molecular gas within Haro 11 for the first time and investigate the gas depletion and star formation processes for the individual starburst regions. 

\subsection{Measurements of molecular gas mass and star formation rate}
\label{subsec:measure_gas_sfr}

We use the CO(1-0) integrated intensity to derive the molecular gas mass, following the relation in \cite{bolatto2013}:
\begin{equation}
M_{\rm H_2} = \alpha_{\rm CO} \times L_{\rm CO}, 
\end{equation}
where the  $M_{\rm H_2}$ is the molecular gas mass in units of solar mass, and the $L_{\rm CO}$ is the luminosity of CO(1-0) emission in units of $\rm K \ \kms \ pc^{-2}$. $L_{\rm CO}$ can be derived by following equation:
\begin{equation}
L_{\rm CO} = 2453 \times I_{\rm CO} \times D_{\rm L}^2 / (1+z),
\end{equation}  
where $I_{\rm CO}$ is the integrated flux in $\rm Jy\ \kms$, $D_{\rm L}$ is the luminosity distance (Mpc), and the redshift ($z$) is adopted as 0.0206467 \citep{Menacho2019}.  The conversion factor $\alpha_{\rm CO}$ is dependent on the gas-phase metallicity \citep[e.g.,][]{bolatto2013,Hunt2014,shi2016a}. \cite{james2013} and \cite{Menacho2021} mentioned the variations in spatially resolved oxygen abundances between electron temperature method and strong emission line diagnoses \citep{Marino2013} in Haro 11. Here, we select the temperature-based metallicities in \cite{james2013} and adopt the area size weighted average metallicity value of $\oh \sim 8.12$. According to the metallicity-$\alpha_{\rm CO}$ relation of \cite{shi2016a}, we assume the average $\alpha_{\rm CO}$ of $60 \ \msun \rm (K \ \kms \ {pc}^{-2})^{-1}$, with a range of $10 - 110 \ \msun \rm (K \ \kms \ {pc}^{-2})^{-1}$. We calculate the gas mass uncertainty by considering the $\alpha_{\rm CO}$ range and the three times of the CO intensity RMS. We then divide the molecular gas mass by the area size to compute the gas mass surface density $\Sigma_{\rm H_2}$. Adopting the CO velocity width as 200 $\kms$, the $\Sigma_{\rm H_2}$ at 4 $\times$ RMS threshold is about $2.3 \times 10^8 \msun \rm \ kpc^{-2}$.

Based on the ``Case B'' recombination model and \cite{Calzetti2000a} reddening formalism, we obtain the extinction-corrected $\ha$ flux using the $\ha/\hb$ ratio. According to the \cite{Kroupa2001} IMF and the calibration relation in \cite{Kennicutt1998}, 
\begin{equation}
{\rm SFR}(\msun {\rm \ yr^{-1}}) = 7.92 \times 10^{-42} \times L_{\rm cor} \rm (\ha)(erg \ s^{-1}),
\end{equation}
the SFR can be determined from $\ha$ luminosity, $L_{\rm cor}(\ha)$. The SFR uncertainty is computed from the $\ha$ luminosity and attenuation uncertainties, which include the $\hb$ intensity errors. Since most of the star formation in galaxies is discrete, \cite{dasilva2014a} reported that different SFR tracers could produce large stochastic fluctuation ($\ge$ 0.5 dex) at the lowest SFR regimes. However, the uncertainty is negligible at a SFR higher than $1 \ \msun\rm \ yr^{-1}$ \citep{dasilva2014a}. Furthermore, the intrinsic SFR might be variable at a shorter time scale in a merging or interacting galaxy. The ionization-based tracers (e.g., $\ha$) will be preferable since we will focus on the intense starburst regions in Haro 11, which have high SFR ($\ge 1 \ \msun\rm \ yr^{-1}$). We compute the SFR surface density, $\Sigma_{\rm SFR}$ taking into account the area of the regions.

\subsection{CO(1-0) distribution and kinematics} 
\label{subsec:co10_distri}

In the $left$ panel of Fig. \ref{fig:Ha_CO10}, we show the integrated intensity map of $\ha$ emission. The star-forming knots (A, B, C, and T) identified in Section \ref{subsec:knots} are marked as gray contours. The corresponding circular regions with equal areas are shown with blue circles. We will use these circle regions in the subsequent analysis. A large fraction ($\sim 44\%$) of star formation rate is occurring within knot B. In the $right$ panel, we present the integrated CO(1-0) intensity (moment 0) distribution contoured by $\ha$ flux. The knots are also marked as blue circles. We note that molecular gas is concentrated around knots B and C, which contain about $55\%$ and $4\%$ of the gas mass, respectively. We also detect an offset between the CO and $\ha$ intensity peaks, indicating the delay between compression of molecular gas and current star formation. The rest of the gas is distributed between the knots B and C. No CO was detected at a significant level toward knots A and T. Furthermore, it seems that the star formation regions locate surrounding the remaining molecular gas, which might indicate two possible scenarios.  One is that we did not detect the molecular gas associated with the recent star formation that traced by $\ha$ emission. Another is that the early-stage star formation occurs within the CO peak, probably surrounded by dusty ISM, thus the $\ha$ emission is not luminous yet.

The systemic velocity ($V_{\rm sys}$) of CO molecular gas in Haro 11 is estimated to be 6051.3 $\pm$ 2.1 \kms. Fig. \ref{fig:CO10_Ha_vel} shows the intensity-weighted velocity field (moment 1, $V_{\rm obs} - V_{\rm sys}$) map of CO(1-0) and $\ha$ emission, as well as the velocity map of stellar components derived from the SSP fitting results. The $\ha$ flux contours are also overlaid on these panels, helping to visualize the spatial association between the molecular gas and star formation regions. The velocity distribution of CO(1-0) is approximatively symmetric to the stellar center. The gas around the eastern region C is approaching, while the western side of knot B shows receding motions. Similarly, the apparent receding motion of the $\ha$ gas, showing as a large \textit{arc} shell shape, is detected at the western side of knot B (around knots A and T) while the gas is approaching around the knot C. However, the stellar components at knots A and C show approaching motions while they are receding at knot B. The irregular stellar velocity is not a circular rotation disk within Haro 11 but a complex merging system. \cite{Menacho2019} studied the slices of the $\ha$ emission in velocity space and compared the ratio maps of $\oiii\lambda5007/\ha$ and $\oi\lambda6300/\ha$.  The authors reported the existence of an expanding shell (\textit{arc} shell shape at the western side of knot B) and the outflows of ionized gas around knot C.  The CO(1-0) also shows the approaching (and receding) motions at the knot C (and B), similar to that of $\ha$, which may indicate the presence of molecular outflows. Furthermore, we note a clear stellar velocity edge between  knots B and C, which probably indicates the collision of stellar components from the galaxy progenitor. Thus, the molecular gas between knots B and C might be the complex/combined stage of collision and feedback.

In Fig. \ref{fig:CO10_Ha_spec}, we present the CO(1-0) and $\ha$ spectra within knots B, C, $r\sim 0.6 \arcsec$ region between B and C (hereafter B/C), as well as the total area. We perform one Gaussian function fitting for the CO(1-0) and $\ha$ spectra in each region. Adopting the systemic CO velocity as $6051.3 \pm 2.1 \kms$, the CO velocity in knot B (C) is red-shifted (blue-shifted), which is consistent with the CO velocity map in Fig. \ref{fig:CO10_Ha_vel}. However, the $\ha$ velocity in knots B and C are blue-shifted, different from the CO velocity. After subtracting the instrument resolution from the fitted Gaussian widths, we compute the velocity dispersion of $\ha$ and CO gas. We note that knot B shows the highest velocity dispersion value of CO and $\ha$ emission, indicating the strongest starburst activity in Haro 11. Furthermore, the velocity dispersion of CO and $\ha$ emission at knot C is also high, which we will discuss in the next paragraph. To clearly understand the molecular gas distribution in Haro 11, we present the CO(1-0) channel maps at widths of 20 \kms in Fig. \ref{fig:CO10_mom1_channels}, contoured on the stellar image from MUSE data. 

\begin{figure*}
   \centering
   \includegraphics[width=0.45\textwidth]{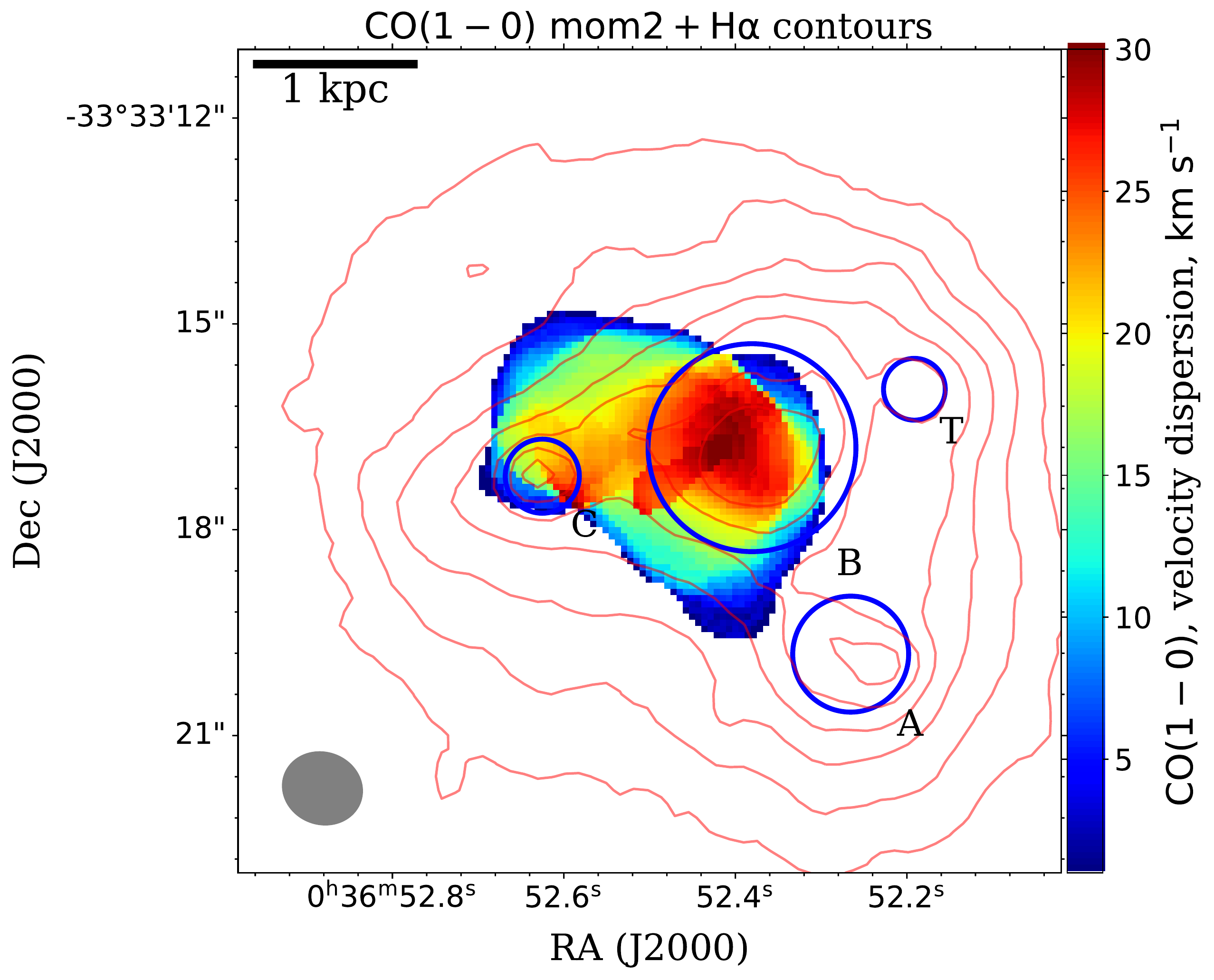}
   \includegraphics[width=0.45\textwidth]{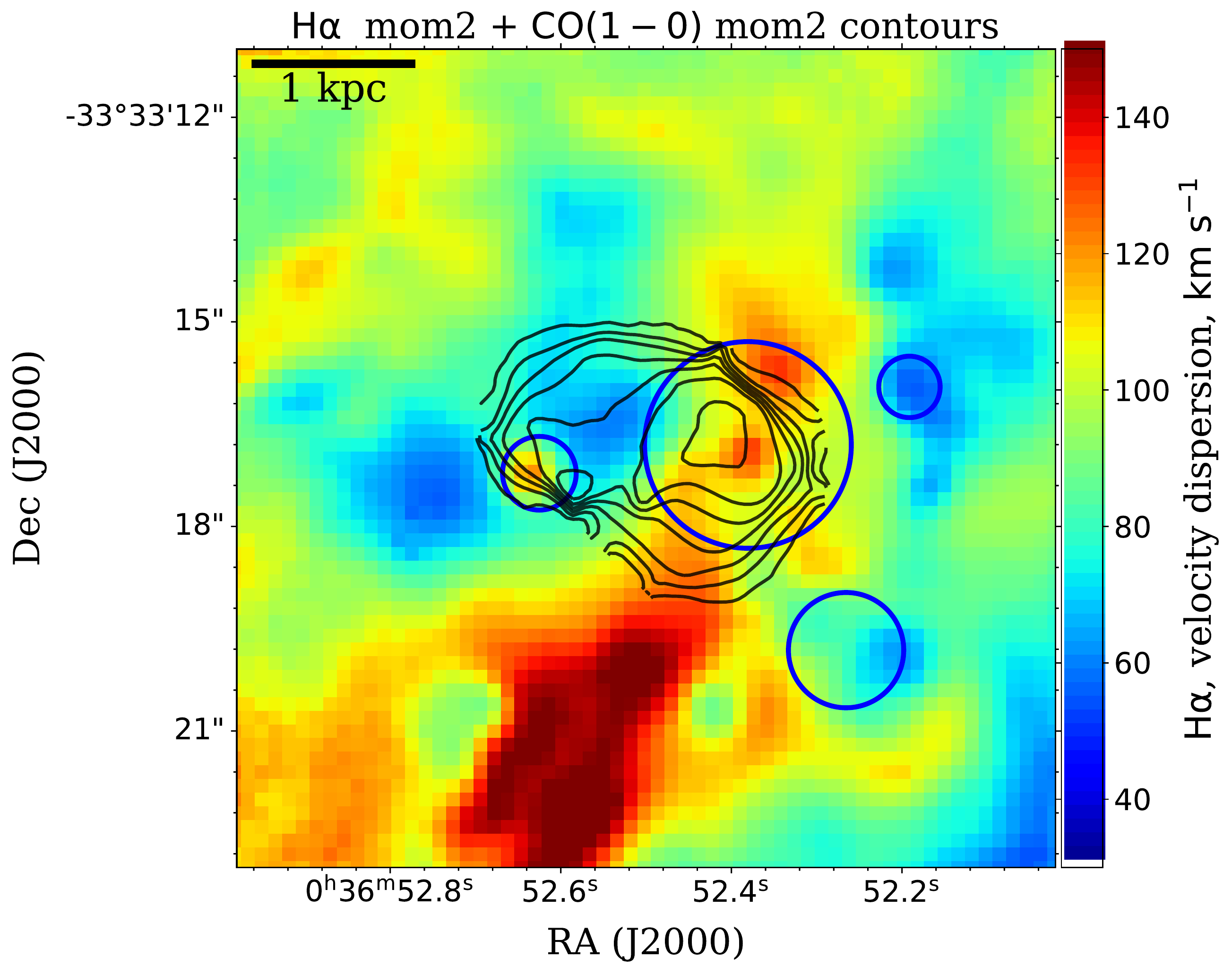}
   \caption{$Left$: The velocity dispersion (moment 2) of CO(1-0) emission line contoured by $\ha$ flux in Haro 11. $Right$: The velocity dispersion of $\ha$ contoured by CO(1-0) velocity dispersion. Other symbols are the same as Fig. \ref{fig:Ha_CO10}. }
   \label{fig:CO10_mom2}
  \end{figure*}

In Fig. \ref{fig:CO10_mom2}, we present the intensity-weighted velocity dispersion (moment 2, $\sigma_v$) maps of CO(1-0) ($left$ panel) and $\ha$ ($right$ panel) emission lines. We note that the CO(1-0) velocity dispersion within/between knots B and C, which contain the high surface density gas, is larger than the relatively diffuse gas in the outer regions. The highest velocity dispersion values of CO gas are detected in the lane extending from knots B toward its southeastern direction. This shape is consistent with the $\nii$-enriched area and then toward the high $\ha$ velocity dispersion region \citep[see Fig. 3 in ][]{Menacho2021}, indicating strong starburst activity, e.g., Wolf-Rayet (WR) stars, are occurring within the collision/overlap region of molecular gas. However, the large velocity dispersion might also be caused by the beam-smearing effect.

\subsection{Spatial star formation efficiencies and specific SFR}
\label{subsec:ssfr_sfe}

   \begin{figure*}
   \centering
   \includegraphics[width=0.45\textwidth]{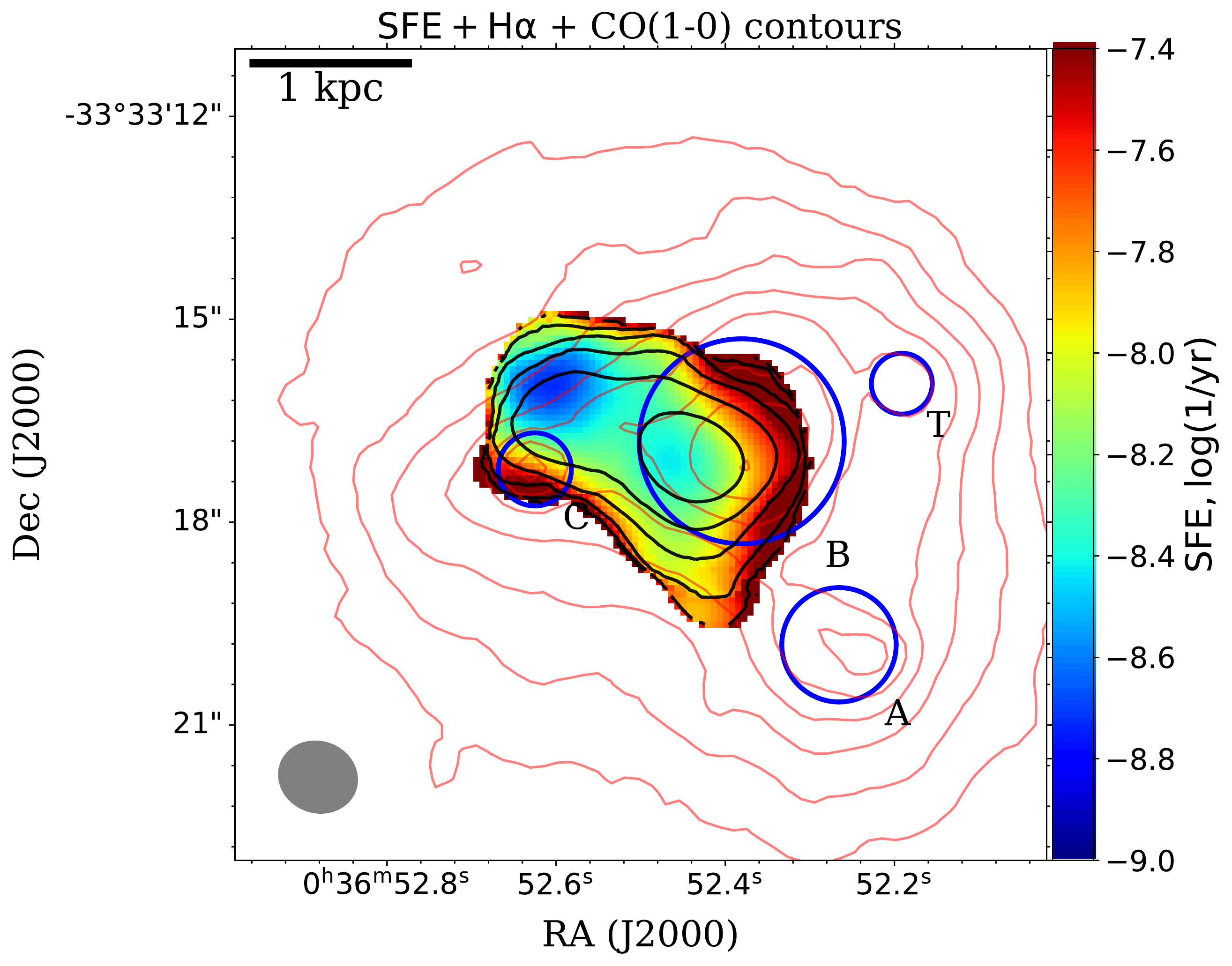}
   \includegraphics[width=0.45\textwidth]{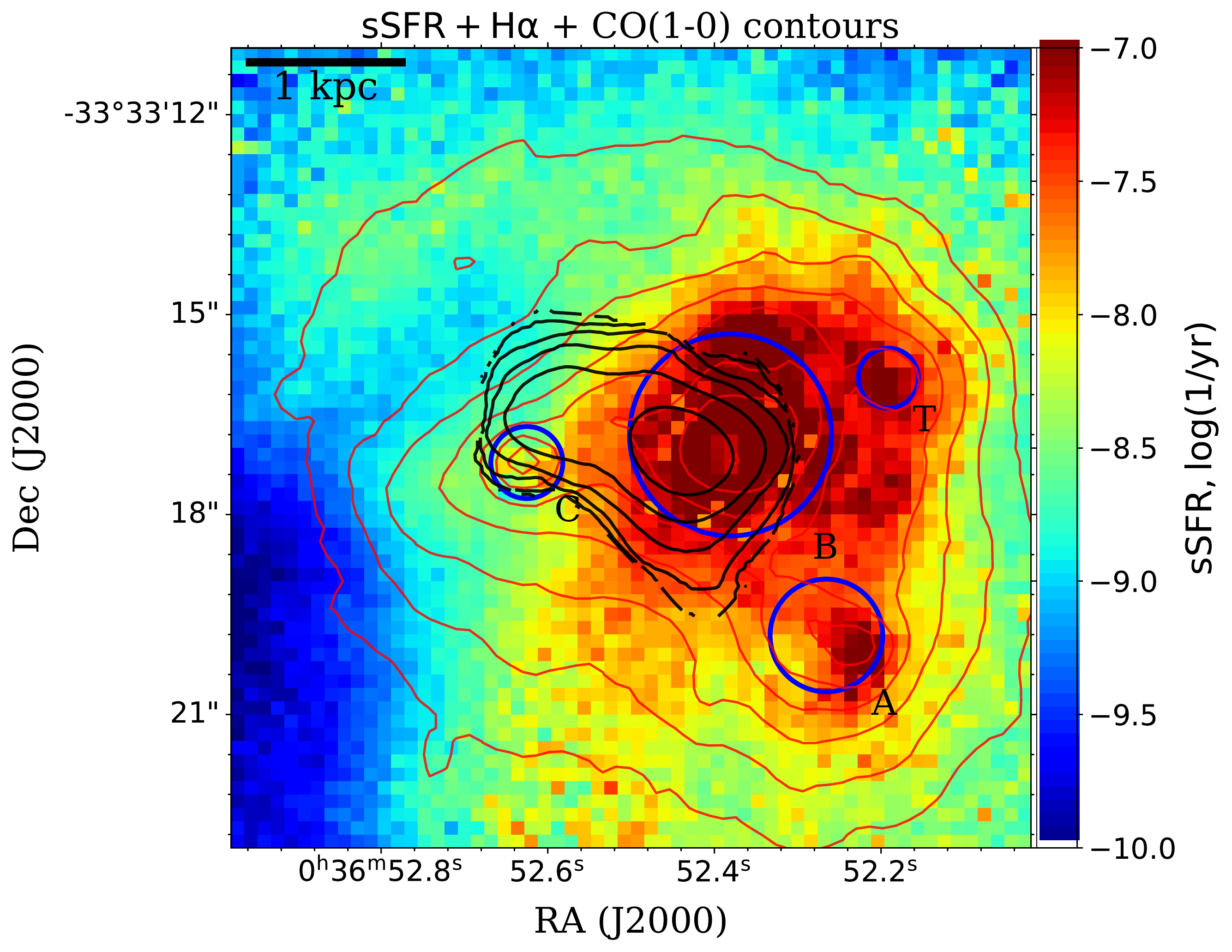}
   \caption{Maps of SFE ($left$ panel) and specific SFR (sSFR, $right$ panel) contoured by $\ha$ (red) and CO(1-0) (black) fluxes. Other symbols are same to Fig. \ref{fig:Ha_CO10}.}
   \label{fig:SFE_sSFR}
   \end{figure*}

To investigate the relative strength of star formation activities, we will discuss the star formation efficiency (SFE) and specific SFR. There are a few different definitions for SFE \citep[e.g., ][]{leroy2008,Kennicutt2012}, but we adopt the ratio of SFR and molecular H$_2$ gas mass as the SFE (SFR/$M_{\rm H_2}$, or $\Sigma_{\rm SFR}/\Sigma_{M_{\rm H_2}}$) throughout this paper. The SFE is the inverse of the molecular gas depletion time, with units of yr$^{-1}$, meaning the time required to consume the molecular gas at the present-day star formation. The specific SFR (sSFR), defined as SFR/$M_*$ or $\Sigma_{\rm SFR}/\Sigma_{M_*}$, represent the stellar mass birthrate in galaxies, will help to characterize the star formation history.

We adopt a Gaussian kernel and use the CASA task \textit{imsmooth} to smooth the original SFR and $M_*$ maps. We set the resolution of target maps as the beam size of CO(1-0) image, then compute the spatial distribution of SFE and specific SFR (sSFR). In Fig. \ref{fig:SFE_sSFR}, we show the SFE and sSFR distribution maps overlaid by $\ha$ and CO(1-0) fluxes. The logarithmic SFE values range from -9.0 to -7.0, which are about 0.2 - 2.0 dex higher than the main sequence of normal star-forming galaxies in the local universe \citep{Kennicutt2012}. The highest SFE value is detected in the northwestern edge of knot B, which consists of the local peaks of $\ha$ emission, but molecular gas surface density is relatively low. The logarithmic sSFR values range from $-9.0$ to $-7.0$, about 1.0 to 3.0 dex above the main sequence of normal star-forming galaxies in the local universe \citep{saintonge2016}. We find the highest sSFRs are detected in knot B, indicating that the stellar mass will assemble efficiently therein. Furthermore, because of the presence of a large amount of molecular with weak $\ha$ emission in the northern and eastern region, toward knots C and B, the sSFRs and SFEs are relatively small. However, the molecular gas will provide the fuel for future star formation around knot B.

\subsection{Global star formation}
\label{subsec:sf_knots}

   \begin{figure*}
   \centering
   \includegraphics[width=0.45\textwidth]{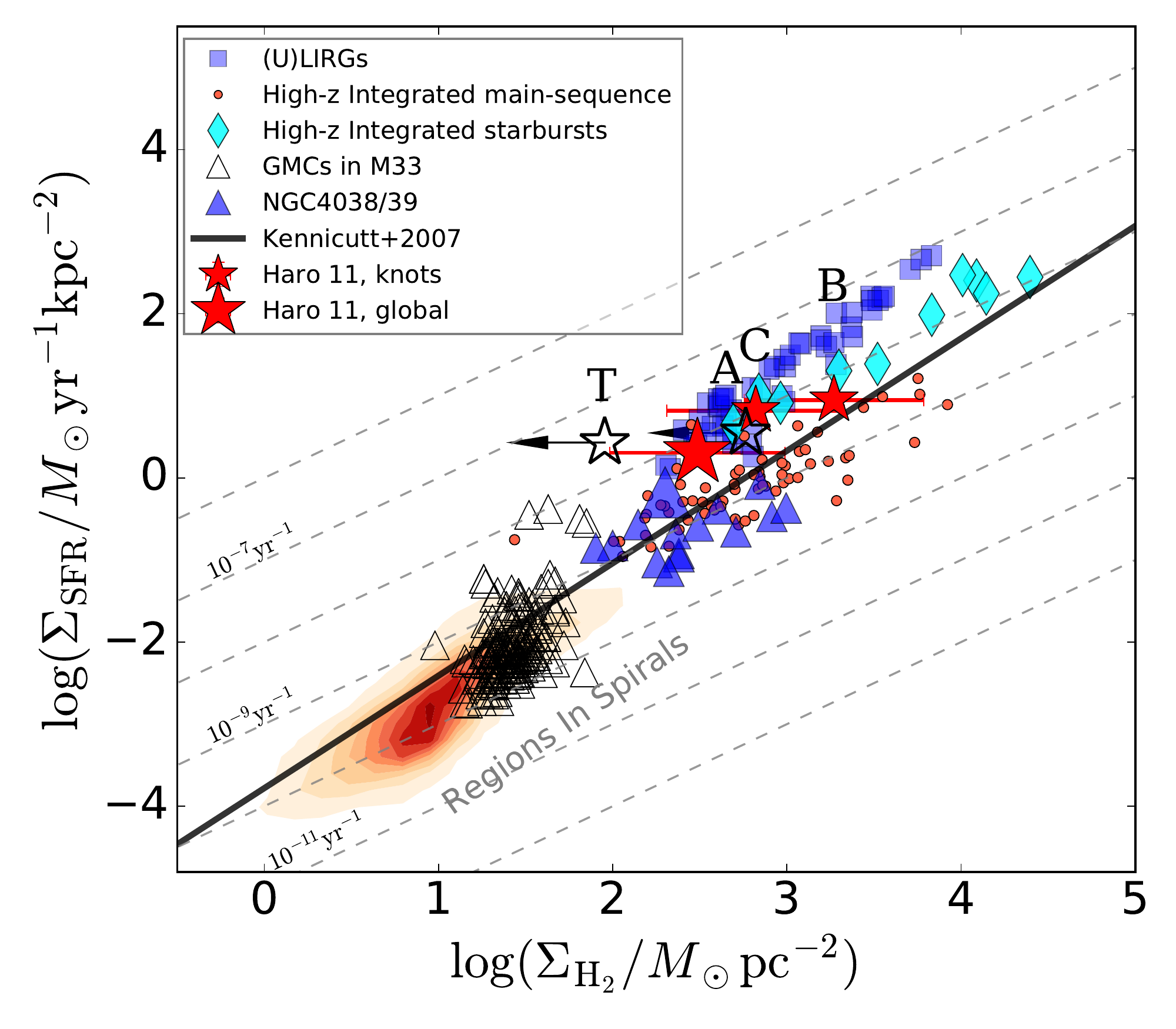}
   \includegraphics[width=0.45\textwidth]{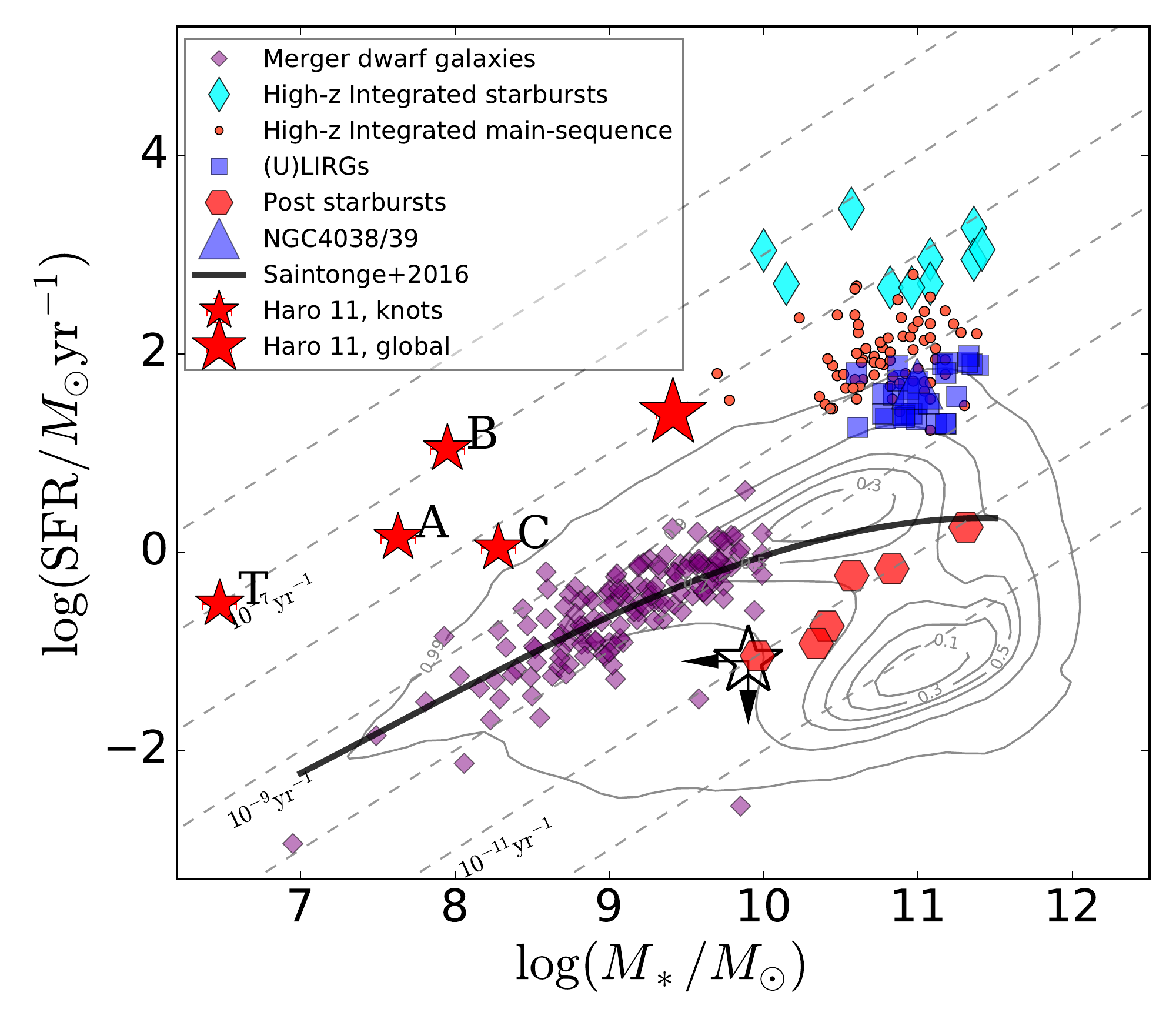}
   \caption{$Left$: Kennicutt-Schmidt (K-S) law, i.e., gas surface density $\Sigma_{\rm H_2}$ vs. SFR surface density $\Sigma_{\rm SFR}$, for different galaxy samples. Red stars mark the knots in Haro 11, and a larger red star means the global value. Because of the lack of CO detection at knots A and T, the upper limited $\Sigma_{\rm H_2}$ are shown as white stars. The star-forming clumps in the nearby merger Antennae (NGC 4038/4039) galaxy are marked as blue triangles \citep{Tsuge2021}, the global value is shown as the larger blue triangle \citep{gao2001, Lahen2017}. The spatially resolved properties in nearby (U)LIRGs from \cite{Wilson2019} are marked as blue squares. Other samples are collected from \cite{Shi2018}, including local spiral galaxies \citep[red contours,][]{bigiel2008,shi2011}, giant molecular clouds in M33 \citep[white triangles,][]{gratier2010}, high-$z$ main-sequence galaxies \citep[red circles: $z \sim 1.2 - 2.2$, ][]{tacconi2013} and starburst galaxies \citep[cyan diamonds: $1.2 < z < 6.3$, see the references in ][]{Shi2018}. The black line represents the K-S law derived by \cite{Kennicutt2007}. The gray dashed lines represent different SFE levels. $Right$: main sequence relation, i.e., stellar mass $M_*$ vs. SFR, for different galaxy samples. Global properties of massive merger (U)LIRGs from \cite{Shangguan2019a} are shown as blue squares. Post starburst galaxies from \cite{smercina2018,smercina2021} are represented with hexagons, in which SFRs are derived from $\ha$ luminosities. The purple diamonds represent the merging dwarf galaxies in the local universe \citep{Paudel2018}, in which the SFRs are estimated with the far-ultraviolet (FUV) fluxes from \textit{GALEX} all-sky survey. Grey contours indicate all of the galaxies derived from the SDSS DR8 MPA-JHU catalog. The main sequence relation of SFGs in the local universe is shown as a solid black line \citep{saintonge2016}. The gray dashed lines represent different sSFR levels. The white star means the post-starburst phase of Haro 11 in the future, where arrows represent the upper limit values of stellar mass and SFR. Other symbols are same to the $left$ panel.}
   \label{fig:KS_law}
   \end{figure*}

In Section \ref{subsec:ssfr_sfe}, we illustrate the spatial distribution of SFE and sSFR. This section will compare the global SFE and SFR of Haro 11 with other galaxies in the literature and explore the possible evolution path of such a late-stage merger system. We calculate the gas mass from CO(1-0) luminisity is about $3.8^{+3.3}_{-3.1} \times 10^9 \ \msun$, in which the uncertainty is caused by the $\alpha_{\rm CO}$ range and the three times of the CO intensity RMS.  In Fig. \ref{fig:KS_law}, we present the Kennicutt-Schmidt (K-S) law and main sequence relation for Haro 11 and other galaxy samples. The data includes individual GMCs in the nearby spiral galaxy M33 \citep{gratier2010}, individual star-forming regions (and global properties) of the merger system NGC 4038/4039 \citep{gao2001,Lahen2017,Tsuge2021}, spatially resolved measurements of 12 spiral galaxies \citep{bigiel2008,shi2011} and nearby merger (U)LIRGs \citep[e.g.,][]{Shangguan2019a,Wilson2019}. In addition, the data also consist of the global properties of post starburst galaxies \citep{smercina2018,smercina2021}, nearby merging dwarf systems \citep{Paudel2018}, high-$z$ main-sequence galaxies \citep[$z \sim 1.2 - 2.2$, ][]{tacconi2013} and starburst galaxies \citep[$1.2 < z < 6.3$, see the references in ][]{Shi2018}. We also overlay the $M_* - \rm SFR$ distribution of all of the galaxies derived from the SDSS DR8 MPA-JHU catalog\footnote{\url{https://www.sdss.org/dr12/spectro/galaxy_mpajhu/}}. The K-S law \citep{Kennicutt2007} and main-sequence relation \citep{saintonge2016} for local galaxies are shown as solid black lines, respectively. The individual knots in Haro 11 are marked as red stars, while the global value is shown as a larger red star.

We find that star-forming regions in the disk of spiral galaxies (including M33), overlap regions of NGC 4038/4039, and high-$z$ normal SFGs follow the well-defined K-S law from \cite{Kennicutt2007}. We find the star formation at knot B is nearly consistent with the K-S law. However, the global SFEs of Haro 11 and knot C are significantly higher ($\sim 0.5$ dex) than these normal SFGs and NGC 4038/4039. The molecular gas depletion time in Haro 11 (and its knots) is similar to the high-$z$ starburst galaxies and star-forming regions in nearby (U)LIRGs. These nearby interacting dwarf galaxies \citep[$M_* < 10^{10} \msun$, $z < 0.02$, ][]{Paudel2018}, selected by visual inspection from both SDSS and DESI Legacy imaging \footnote{\url{https://www.legacysurvey.org/}} surveys, are consistent with the star-forming main-sequence relation \citep{saintonge2016}. Most of these galaxies are located at the pre-merger or early-merger stage, suggesting that the significant enhancement of star formation has not been triggered yet. The sSFRs ($\geq 10^{-8} \ \rm yr^{-1}$) of Haro 11 and its knots are resemble to the high-$z$ starburst galaxies, which are about 0.6 - 2 dex higher than other galaxies. Adopting the $\ha$-based SFRs, \cite{smercina2018,smercina2021} locates these post-starburst galaxies at the transition phase between major gas-rich mergers and gas-poor quiescent galaxies, where the star formation is suppressed by continued injection of turbulent or mechanical heating.

Considering the similar SFEs and sSFRs as high-$z$ starburst galaxies, as well as the small stellar mass and low metallicity (1/3 $Z_{\odot}$), we can regard Haro 11 as an analog of high-$z$ dwarf starburst galaxies. We will explore the molecular and neutral gas depletion and discuss the evolution path in Section \ref{subsec:hi}. 

\subsection{$\ci$ distribution}
\label{subsec:ci}

   \begin{figure*}
   \centering
   \includegraphics[width=0.45\textwidth]{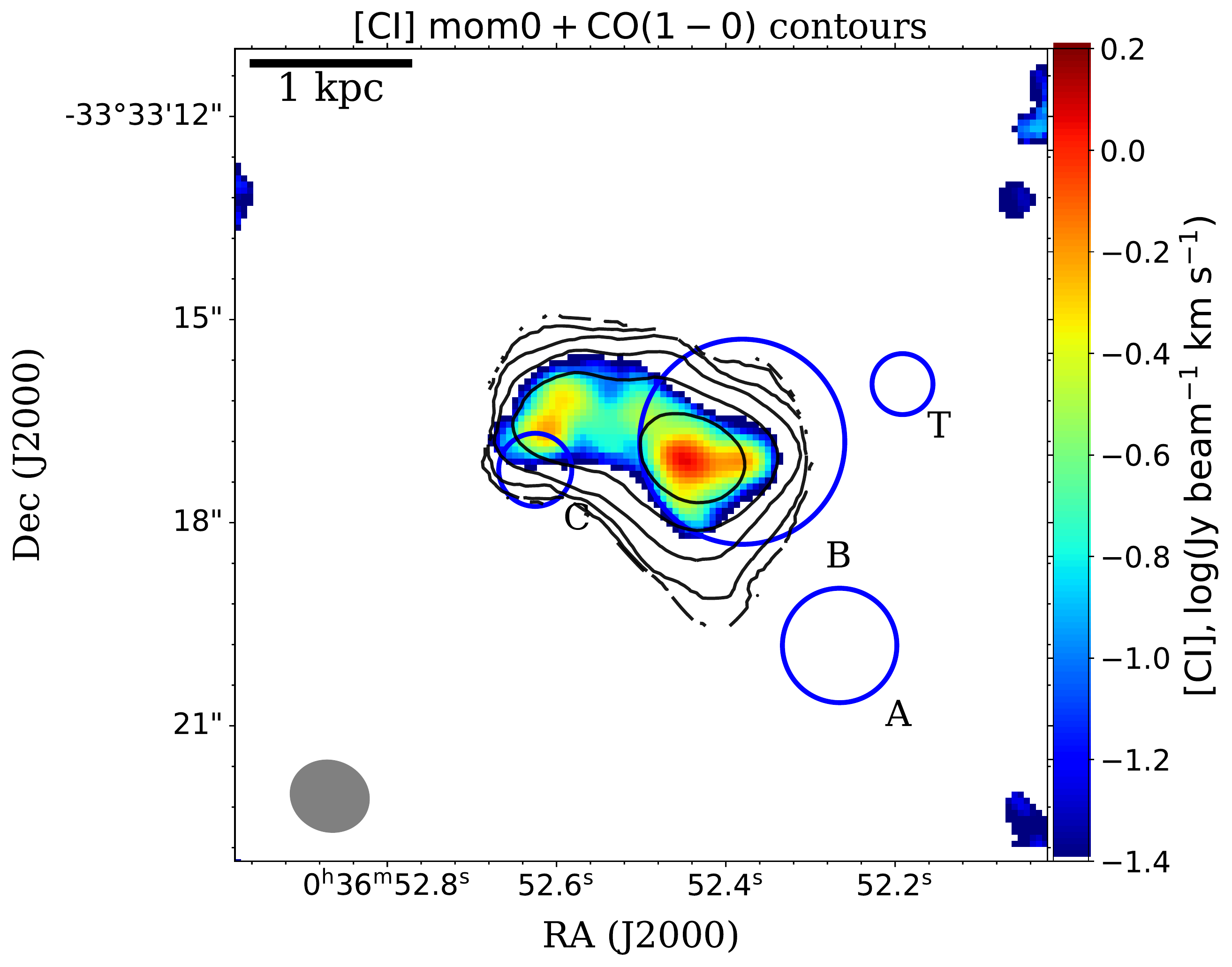}
   \includegraphics[width=0.45\textwidth]{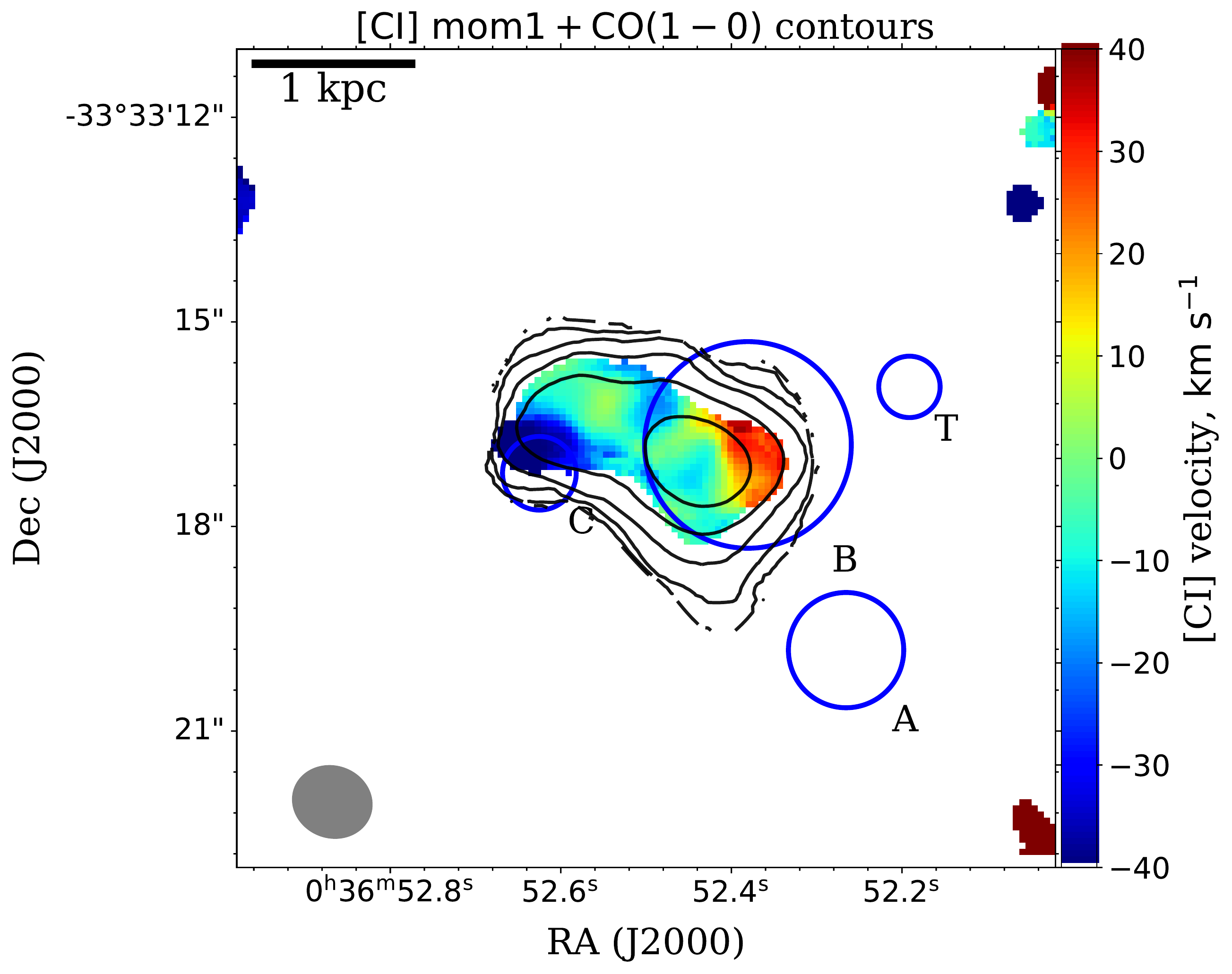}
   \includegraphics[width=0.45\textwidth]{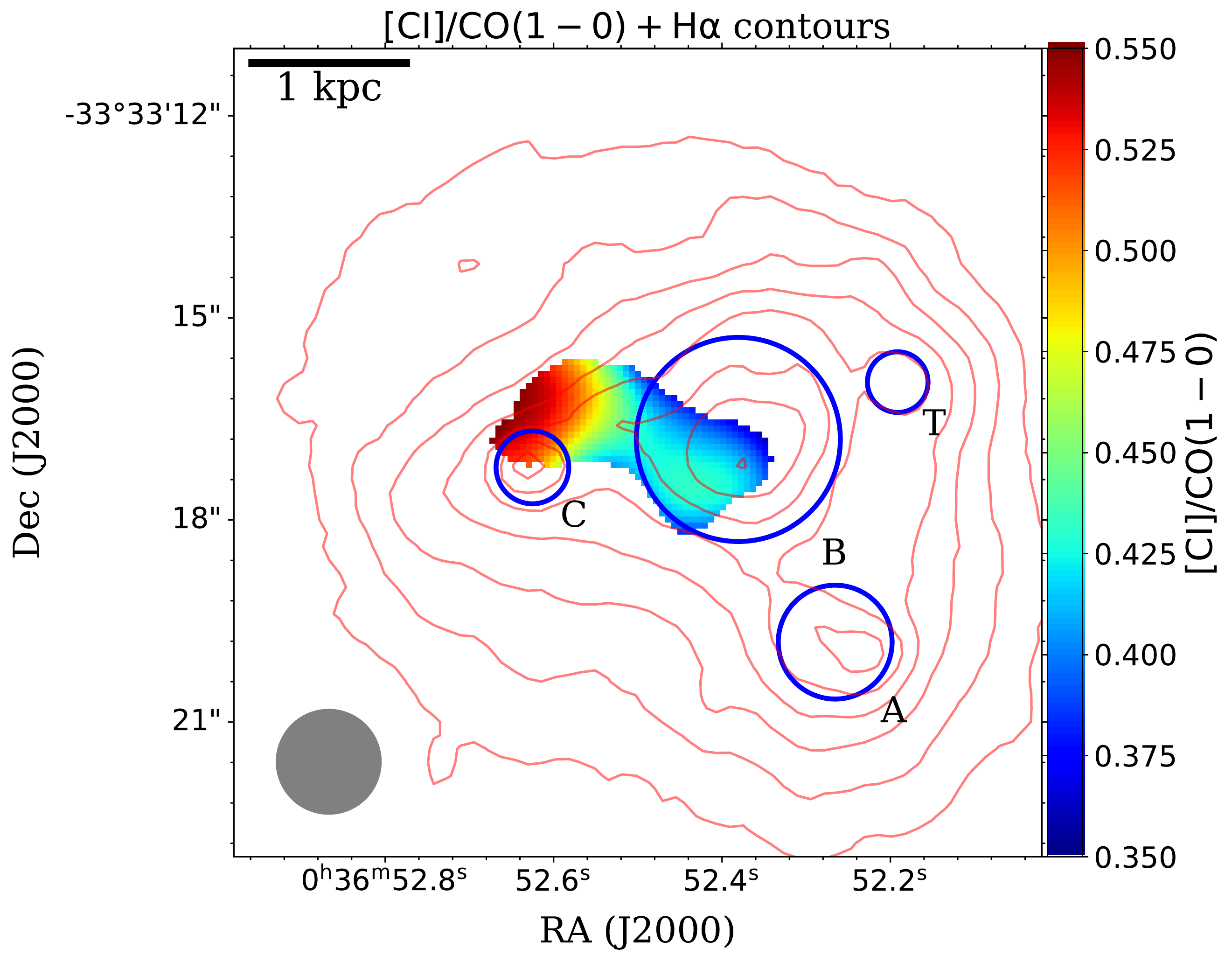}
   \includegraphics[width=0.45\textwidth]{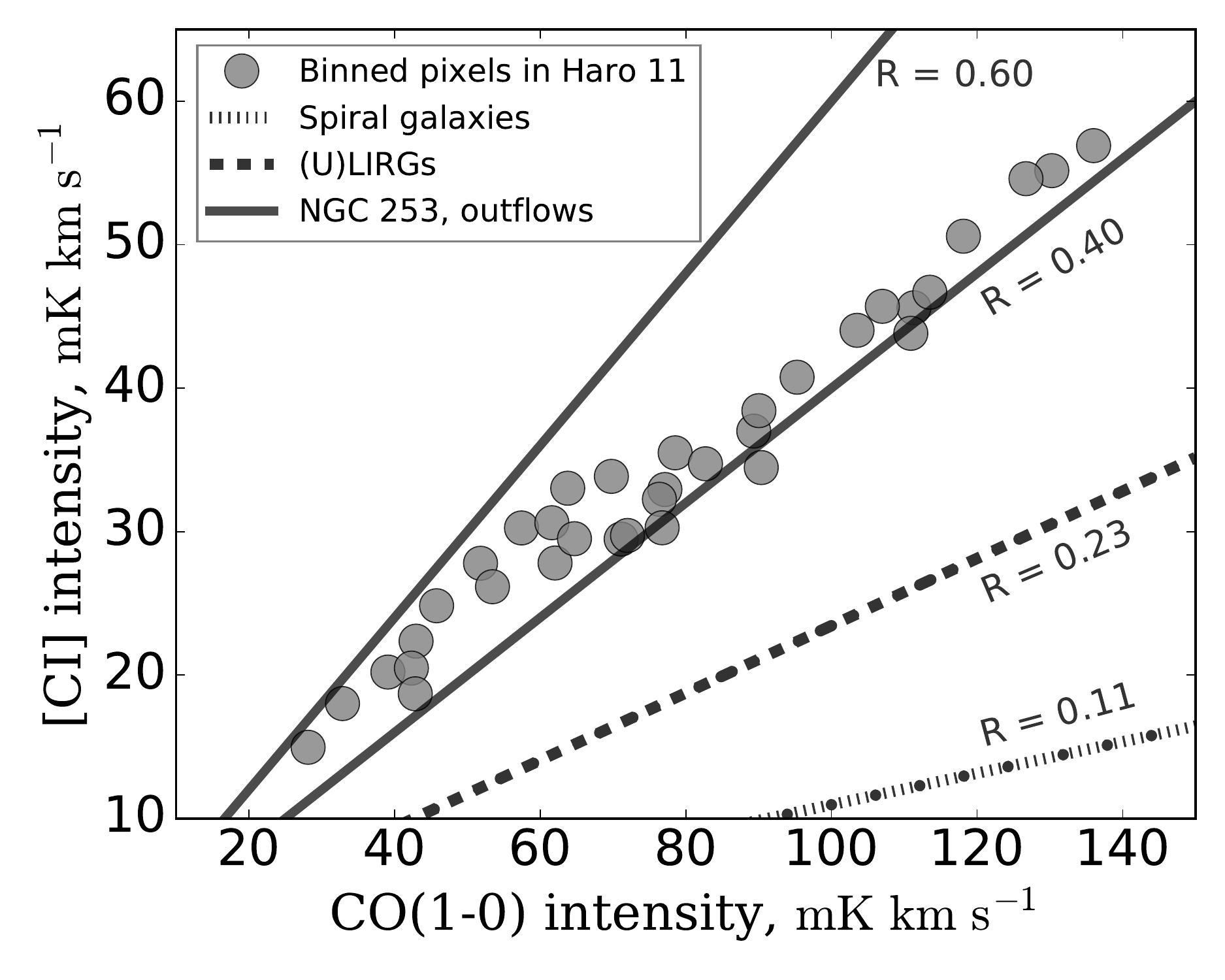}
   \caption{$Upper$: Distribution of $\ci$(1-0) emission (color) intensity ($left$) and velocity ($right$) contoured with the CO(1-0) emission (black contours) of Haro 11, with levels of $(1/2)^n$ ($n \ =$ 1, 2, ..., 8) times of the maximum value of CO intensities. $Bottom-left$: Distribution of the line ratio ($R = I_{\ci} / I_{\rm CO(1-0)}$) of the [CI] and CO emission intensity. Other symbols are same to Fig. \ref{fig:Ha_CO10}. $Bottom-right$: The integrated intensity of $\ci$ pixels are plotted against the CO(1-0), with units of $\rm mK \ \kms$. The pixels in Haro 11 are rebinned as $\sim 0.4\arcsec$. The average ratios in the disks of spiral galaxies and in (U)LIRGs reported by \cite{Jiao2019} are shown as dotted ($R \sim 0.11$) and dashed ($R \sim 0.23$) lines, respectively. Black solid lines represent the ratios ($R \sim 0.4 - 0.6$) in the outflow regions that emerge from the nucleus of NGC 253 \citep{Krips2016}.}
   \label{fig:CI_CO10}
   \end{figure*}

Though the CO emission line is commonly used as the molecular gas tracer in ISM, its reliability is dependent on the optical depth, metallicity, and radiation strength. Previous observations \citep[e.g., ][]{ikeda2002,shimajiri2013} find that neutral carbon is well mixed throughout the clouds, while not only locating at the layer interface between the ionized carbon regions and the shielded CO molecule cores. Furthermore, the integrated intensity ratios ($R = I_{\ci} / I_{\rm CO(1-0)}$) are significantly higher in active nuclear environments than in galactic clouds. Cosmic rays from supernovae, ultraviolet (UV) radiation from starburst, and the mechanical shocks will dissociate the CO molecules, significantly enhancing the C abundance \citep[e.g.,][]{tanaka2011, Krips2016,bisbas2017,papadopoulos2018}. These findings suggest that neutral carbon can be a complementary molecular gas tracer to CO, especially in active galaxies. 

This section will study the spatially resolved [CI] line emission with ALMA band 8 observations and the line ratios of [CI] to CO in Haro 11, then evaluate its potential as a complementary H$_2$ tracer. We smooth the CO and [CI] brightness images to a larger beam size of 1.5$\arcsec$ $\times$ 1.5$\arcsec$ ($\sim$ 630 $\times$ 630 pc), with the same pixel size of 0.1$\arcsec$. In Fig. \ref{fig:CI_CO10}, we provide the spatial distribution of the CO(1-0) and [CI] emission intensity and velocity, as well as their line intensity ratios. We note the [CI](1-0) line emission is detected throughout the bridge from knot B to C, in which the peak location and distribution strongly resemble the CO line emission. We also check the velocity distribution of [CI](1-0), which presents a similar trend to the velocity of CO(1-0) shown in Fig. \ref{fig:CO10_Ha_vel}. These results seem to suggest  that [CI] can trace the same molecular gas as CO in such a dwarf merger starburst galaxy. Previous researches \citep[e.g.,][]{Israel2015, Jiao2019} have reported the scenario that [CI] luminosity is associated with the CO(1-0) or CO(2-1) emission, suggesting that the [CI] can trace the molecular gas in $\hii$, Seyfert, and starburst galaxies on kpc scales. Using the $\ci$ to H$_2$ conversion calibration of \cite[eq.1 and Fig.6, ][]{madden2020}, we estimate the H$_2$ gas mass to be $\sim 5.3 \times 10^8 \msun$  and $\sim 2.1 \times 10^8 \msun$ (see Table \ref{table:1}) at the metallicity of 0.1 $Z_\odot$ and 0.25 $Z_\odot$, respectively. The calculation uncertainty is about 0.3 dex. The $\ci$-based H$_2$ gas mass is much smaller than the median value of CO-based H$_2$ gas mass, but is nearly consistent with the gas mass range considering the calibration uncertainty. This difference may be due to the larger $\alpha_{\rm CO}$ we adopted or the amount of missing flux in the $\ci$ observation. Further studies are needed to identify what limitations there may be for $\ci$ tracing H$_2$ and how $\ci$ emission translates to H$_2$ mass. Furthermore, in the $bottom-right$ panel of Fig. \ref{fig:CI_CO10}, we compare the $\ci$/CO(1-0) ratios in Haro 11 with the disks of spiral galaxies \citep[$R \sim 0.11$, ][]{Jiao2019}, (U)LIRGs \citep[$R \sim 0.23$, ][]{jiao2017,Jiao2019}, and the nuclear outflow region of the nearby starburst galaxy NGC 253 \citep[$R\sim 0.4 - 0.6$, ][]{Krips2016}. We rebinned the pixels in $\ci$ and CO(1-0) intensity maps as 0.4$\arcsec$. Though the pixel size is smaller than the beam sizes of CO and $\ci$, it is instructive to perform such a comparison. We note the line ratios of [CI]/CO range from 0.35-0.55 within the two local peak molecular gas components, which are significantly enhanced concerning the values of normal spiral galaxies and (U)LIRGs while being similar to the outflow region of NGC 253. \cite{Krips2016} supposed the high ratios in NGC 253 indicate that shocks, associated radiation, and/or cosmic rays affect the carbon/CO excitation and their abundance.  Meanwhile, the line ratios at the north region of knot C regions are larger than 0.45, which are systemically higher than in the central knot B. To explain this [CI] enrichment, we need to remember that Haro 11 is one of the most extreme starburst galaxies in the nearby universe. \cite{hayes2007} detected strong Ly$\alpha$ emission extending from knot C to its north and south sides, probably indicating a possible outflow. Recently, using the spectroscopy data from the HST/COS telescope/instrument, \cite{Oestlin2021} found knot C has the lowest covering fraction ($\leq 50\%$) of neutral gas and the highest Ly$\alpha$ escape fraction. Their results suggested that knot C is the original region of the escaping Lyman continuum photons. 
The intense radiations, such as cosmic rays and far-UV photons, dissociate the CO molecules and enhance the neutral carbon abundance. The higher $\ci$/CO(1-0) ratios may indicate the significant amount of molecular gas not traced by the CO emission, i.e., CO-dark gas, contributing to the larger CO-to-H$_2$ conversion factor in such a metal-poor starburst galaxy \citep[e.g., ][]{bolatto2013}.

\section{Discussion}
\label{sec:dis}

\subsection{Molecular gas in knots A and T}
\label{subsec:CO_AD}

The exciting finding is the relatively high sSFRs within knots A and T, with no significant CO detections, limited by the ALMA detection sensitivity. Considering the high SFR values, the average SFEs within regions A and T might be more prominent than in knots B and C ($left$ panel of Fig. \ref{fig:SFE_sSFR}). Analyzing the $\heii\lambda4686$ features, \cite{james2013} reported that knot A contains about 900 young ($\leq$ 5 Myr) WR stars and presents typical nitrogen abundance. They suggested that the WR phase in knot A is ongoing while the ejected nitrogen-rich material does not have enough time to cool down and to mix with the warm ISM. Knot T is located toward one of the lobes of the dusty arm extending from A to T \citep{Menacho2019,Menacho2021}. \cite{Menacho2019} found the dusty arm shows an arc resembling an expanding ionized gas shell with a velocity of about 50 $\kms$. The compression of molecular gas might trigger the star formation within knot T. The molecular gas within knots A and T has been partly consumed by the intense star-formation or destroyed by the strong existing radiation and shocks. Much deeper observations with the ALMA telescope are needed to investigate the accurate mass distribution and kinematics of molecular gas.

\subsection{Molecular gas properties in previous studies}
\label{subsec:comp_dwarf}

To investigate gas and dust properties in low metallicity environments, \cite{madden2013} performed an extensive program, named ``The Dwarf Galaxy Survey'' (DGS). This program observed about 50 dwarf metal-poor galaxies at the far-infrared (FIR) at sub-millimeter bands with the \textit{Herschel Space Observatory}. The metallicities of these nearby DGS galaxies range from $\sim 2.5\% \ Z_{\odot}$ (I Zw 18) to $\sim 1 \ Z_{\odot}$ (He 2-10). Haro 11 is also included by the DGS, showing the highest global SFR \citep{madden2013,cormier2019}.

The molecular gas tracer CO(1-0) in Haro 11 was not detected by the Mopra telescope \citep{Cormier2014}. However, two other tracers CO(2-1) and CO(3-2) lines are observed with the Atacama Pathfinder EXperiment (APEX). The beam sizes for CO(2-1) and CO(3-2) observations are about 26$\arcsec$ and 17$\arcsec$, respectively, which are much larger than the ALMA observation in this work. To estimate the H$_2$ gas mass, \cite{Cormier2014} adopted two different CO-to-H$_2$ conversion factors, one is the Galactic value ($X_{\rm CO} = 2.0 * 10^{20}\ \rm cm^{-2} (K \ km \ s^{-1})^{-1}$, or $\alpha_{\rm CO} = 4.3 \ \msun \ \rm (K \ \kms \ pc^{-2})^{-1}$ \cite{bolatto2013}), while another is ten times that of the Galactic value. To convert the CO(2-1) and CO(3-2) intensities to CO(1-0) intensity, \cite{Cormier2014} adopted the CO(2-1)/CO(1-0) ratio ($R_{21}$) as 0.8 and CO(3-2)/CO(1-0) ratio ($R_{31}$) as 0.6 \citep[e.g.,][]{leroy2008,wilson2009}. Based on these two assumptions, the H$_2$ gas mass in Haro 11 is estimated as $2.5 \times 10^8 \ \msun$ and $2.5 \times 10^9 \ \msun$. The former value is smaller than our H$_2$ gas mass estimation ($3.8^{+3.3}_{-3.1} \times 10^9\ \msun$) in this work, because the conversion factor of (1 times of) Galactic value would underestimate the molecular gas mass in low-metallicity galaxies \citep{shi2016a, madden2020}. Meanwhile, the later one ($2.5 \times 10^9 \ \msun$) is consistent with our estimation. Furthermore, \cite{Cormier2014} also provide the total molecular gas mass derived from the dust measurements from \textit{Herschel} observations \citep{remy-ruyer2013,remy-ruyer2015}. \cite{remy-ruyer2015} performed a full dust SED modeling for photometric measurements up to 500 $\mu \rm m$, and calculated the dust mass in Haro 11 about $9.9 \times 10^6 \ \msun$. Adopting the dust-to-gas ratio ($D/G$) as 1/150 at a metallicity of 8.7, \cite{Cormier2014} estimated the H$_2$ gas mass as $3.6 \times 10^9\ \msun$. Our result ($3.8^{+3.3}_{-3.1} \times 10^9\ \msun$) is consistent with the dust-based H$_2$ estimation of \cite{Cormier2014}, which suggests that the assumption of $\alpha_{\rm CO}$ in this work is appropriate considering the probable CO-dark molecular gas in low-metallicity galaxies \citep[e.g.,][]{madden1997,madden2000,wolfire2010,madden2020}.

\cite{Cormier2014} found the SFR surface density of Haro 11 is about 2 dex above the K-S relationship, given the low gas mass surface density using CO to trace the total H$_2$ gas mass.
However, in Fig. \ref{fig:KS_law}, our ALMA observations put  Haro 11 closer to the K-S relationship, less than 0.5 dex above, consistent with the finding in \cite{madden2020} that star-forming galaxies fall close to or on the K-S law when taking the CO-dark gas into account. Given the similar H$_2$ gas mass estimations, the difference of the position in the gas mass surface density - SFR surface density diagram is caused by the different estimations of galaxy size. \cite{Cormier2014} used the area size about 1266 kpc$^2$ derived from the \textit{Herschel} photometry apertures \citep{remy-ruyer2013}, while we adopt the area size about 13 kpc$^2$ calculated from the radius $R_{80}$ in $r$ band image. The different aperture sizes cause this difference. The circular aperture size of Haro 11 in the FIR band is about 45$\arcsec$ \citep{remy-ruyer2013}, which determines the morphology properties traced by the dust emission. However, using the high spatial resolution image from MUSE, we find the $R_{80}$ is about 5$\arcsec$, which covers all of the observed CO emission by ALMA and more than 95$\%$ of the $\ha$ emission from MUSE. The area size of Haro 11 in our work focuses on the area around the brightest star-forming regions.  Using different area sizes, Haro 11 will show an offset in the gas mass surface density - SFR surface density diagram. However, the total gas depletion time in this work is still consistent with \cite{Cormier2014}.

\subsection{Turbulent pressure and viral parameter}
\label{subsec:turb_viral}

Though showing similar morphology and kinematics to the merger galaxy NGC 4038/4039 \citep{Oestlin2015}, we note the SFE and sSFR values of star-forming regions in Haro 11 are much larger than those in NGC 4038/4039 and some (U)LIRGs. The gas surface densities of knots B (C) are about $1860 \ (660) \ \msun \rm \ pc^{-2}$, which are systemically larger than the overlap and spiral regions in NGC 4038/4039, supporting the probable explanation that molecular gas is much denser in Haro 11 \citep{Oestlin2015}. Furthermore, we estimate the turbulent pressure ($P_{\rm turb}$) and viral parameter ($\alpha_{\rm vir}$) to investigate why there are different SFEs within star-forming regions in these galaxies.

The viral parameter in a molecular cloud describes the balance between its kinetic energy and gravitational potential \citep[e.g.,][]{Krumholz2005,Kauffmann2013,Sun2018a}. Molecular cloud fragments are supercritical, unstable and tend to collapse when $\alpha_{\rm vir} < 2$, while the gas motion therein may prevent fragments from collapsing if $\alpha_{\rm vir} > 2$. Viral parameter is important in determining the possibility of whether or not the clouds can form stars (clusters) However, except for self-gravity, the magnetic field and external gravitational potential also regulate the cloud dynamics in some environments, such as galaxy outer regions with low gas density and the galactic center with high ambient pressure \citep[e.g.,][]{heyer2001,oka2001,Sun2018a,smercina2021}. If assuming pressure equilibrium within a gas cloud, the internal turbulent pressure (i.e., internal kinetic energy density) will equal the external pressure, which has been emphasized in previous studies of galactic clouds \citep[e.g.,][]{field2011,barnes2020,krieger2020}.  Giant molecular clouds in the Milky Way have typical sizes of radius ranging from 5 to 200 pc \citep[e.g.,][]{murray2011}. The radius of the synthesized beam in our CO(1-0) observation of Haro 11 is about 220 pc, slightly larger than the range of Galactic molecular cloud sizes. We will assume that the beam size represents the relevant cloud size scale in the following analysis of the viral parameter and turbulent pressure.

  \begin{figure}
   \centering
   \includegraphics[width=0.47\textwidth]{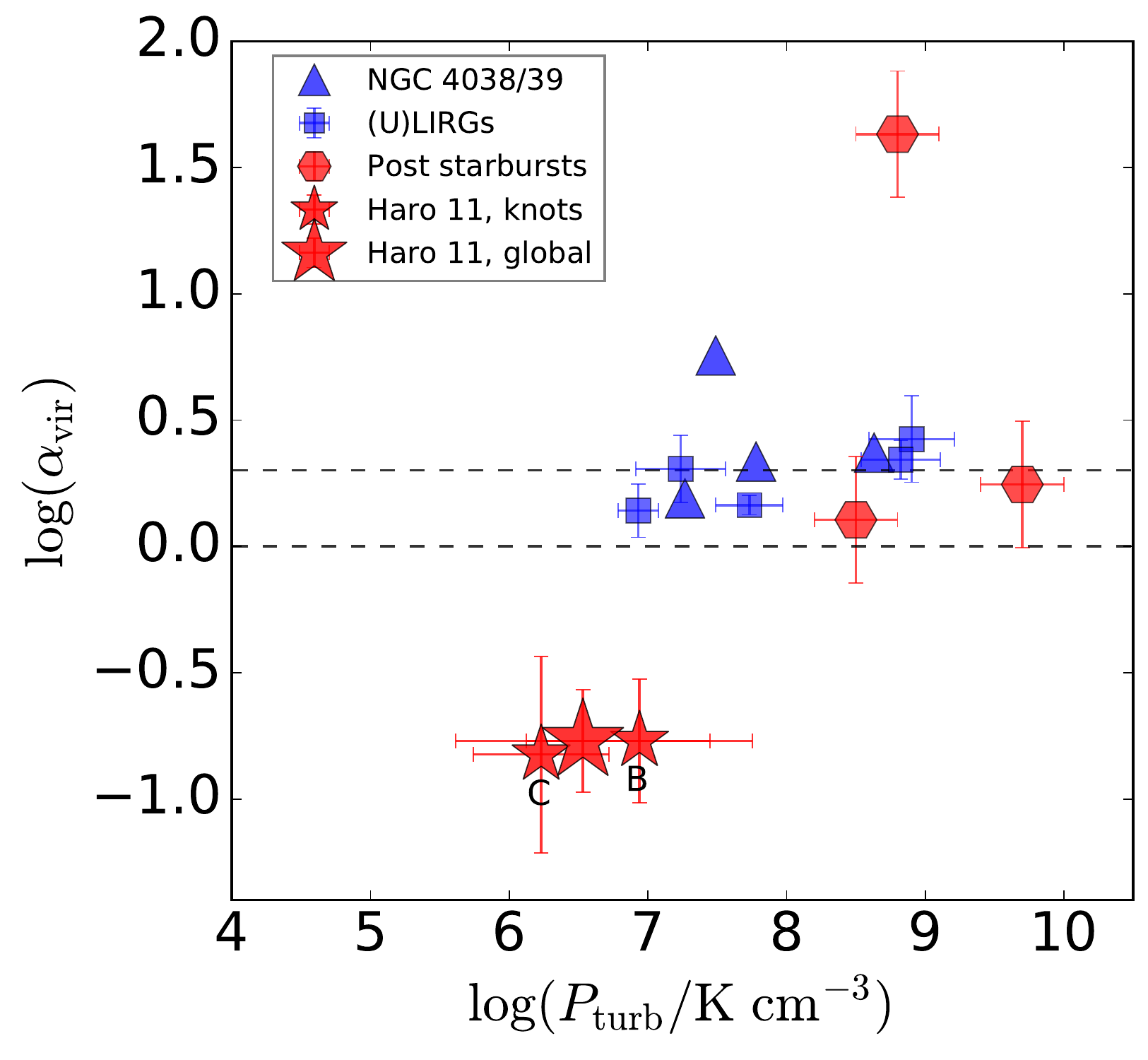}
   \caption{Comparison of the turbulent pressure ($P_{\rm turb}$) and the viral parameter ($\alpha_{\rm vir}$) between Haro 11 and other types galaxies. The properties of knots B, C, and Haro 11 are shown as red stars. Blue squares represent the global properties of five (U)LIRGs calculated from the pixel-binned measurements of \cite{Wilson2019}. The global properties of three post-starburst galaxies derived by \cite{smercina2021} are shown as red hexagons. Blue triangles represent the four young massive clusters forming regions located at the overlap region of NGC 4038/4039 \citep{Tsuge2021}. These errorbars cover the $16 - 84\%$ widths of the distribution of $P_{\rm turb}$ and $\alpha_{\rm vir}$ across all resolved regions in each galaxy. Dashed lines represent the levels of $\alpha_{\rm vir} = 1, 2$, respectively.}
   \label{fig:viral_tub}
  \end{figure}

Following \cite{Sun2018a}, we estimate the $P_{\rm turb}$ and $\alpha_{\rm vir}$ at the scale of the beam size through the following equations:
\begin{equation}
\begin{split}
\alpha_{\rm vir} \simeq \frac{5 \sigma^2 r_{\rm beam}}{f G M} = \frac{5 \rm {ln2}}{\pi f G} (\frac{\sigma_v}{\rm \kms})^2
(\frac{\Sigma_{\rm tot}}{M_{\odot} \ \rm pc^{-2}})^{-1} (\frac{r_{\rm beam}}{\rm pc})^{-1},
\end{split}
\end{equation}
\begin{equation}
\begin{split}
P_{\rm turb}/k_{B} \simeq 61.3 {\rm \ K \ cm^{-3}} (\frac{\Sigma_{\rm H_2}}{M_{\odot} \ \rm pc^{-2}})(\frac{\sigma_{v}}{\rm \kms})^2(\frac{r_{\rm beam}}{\rm 40 \ pc})^{-1}.
\end{split}
\end{equation}
In these equations, the factor $f$ is about 10/9 based on a density profile $\rho(r) \propto r^{-1}$ \citep{Sun2018a}. The gravitational constant $G \simeq 4.3 \times 10 ^{-3} \ {\rm pc} \ \msun^{-1} \ (\rm \kms)^2$, and $r_{\rm beam}$ is the radius of the synthesized beam. The $\Sigma_{\rm tot}$ is the surface density of total mass, including the stellar mass and gas mass. In Fig. \ref{fig:viral_tub}, we perform a comparison of the turbulent pressure and the viral parameter between Haro 11, (U)LIRGs, NGC 4038/4039, and post starburst galaxies. The values of $\alpha_{\rm vir} = 1, 2$ are marked as dashed lines.  We note the $\alpha_{\rm vir}$ and $P_{\rm turb}$ of knots B and C are significantly smaller than those of the clumps in massive merger system NGC 4038/4039, which probably lead to the higher SFE values in the starburst regions of Haro 11. Meanwhile, the median values of $\alpha_{\rm vir}$ and $P_{\rm turb}$ of Haro 11 are 1 - 4 orders of magnitude lower than the median values measured for star-forming regions of (U)LIRGs \citep{Wilson2019} and post starburst galaxies \citep{smercina2021}.  (U)LIRGs and post-starburst galaxies show larger $\alpha_{\rm vir}$ and $P_{\rm turb}$ values, supporting the negative feedback scenario, such that energy injection from AGN and stellar feedback will dissipate and heat the molecular gas, suppress the star formation, and accelerate the quenching of star formation in the host galaxy. \cite{james2013} studied the $\heii\lambda4686$ features and revealed large young WR populations ($> 1000$ stars) in knots A and B. In addition, bright X-ray emission is detected at and around knots B and C \citep[e.g.,][]{grimes2007,Gross2021}, possibly originating from two intermediate-mass black holes (IMBHs), with masses of about $M_{\bullet} \gtrsim 7600 \ \msun$ and $M_{\bullet} \gtrsim 20 \ \msun$, respectively. Due to the complex kinematics of gas and stars in the center of the merger system, the IMBH therein might accrete the material rapidly and evolve into a supermassive black hole. The mechanical feedback from a massive black hole or the starburst, such as jets and outflows, will photoionize the molecular gas and keep the ISM warm or entrain them to escape from the host galaxy, thus suppressing star formation.
Considering stellar feedback or the potential AGN activities in Haro 11, the turbulent pressure and the viral parameter probably follow a similar evolution trend as (U)LIRGs or post-starbursts, leading to a quenching timescale shorter than the molecular gas depletion time (1/SFE, $<$ 0.15 Gyr).

\begin{figure*}[!t]
   \centering
   \includegraphics[width=0.45\textwidth]{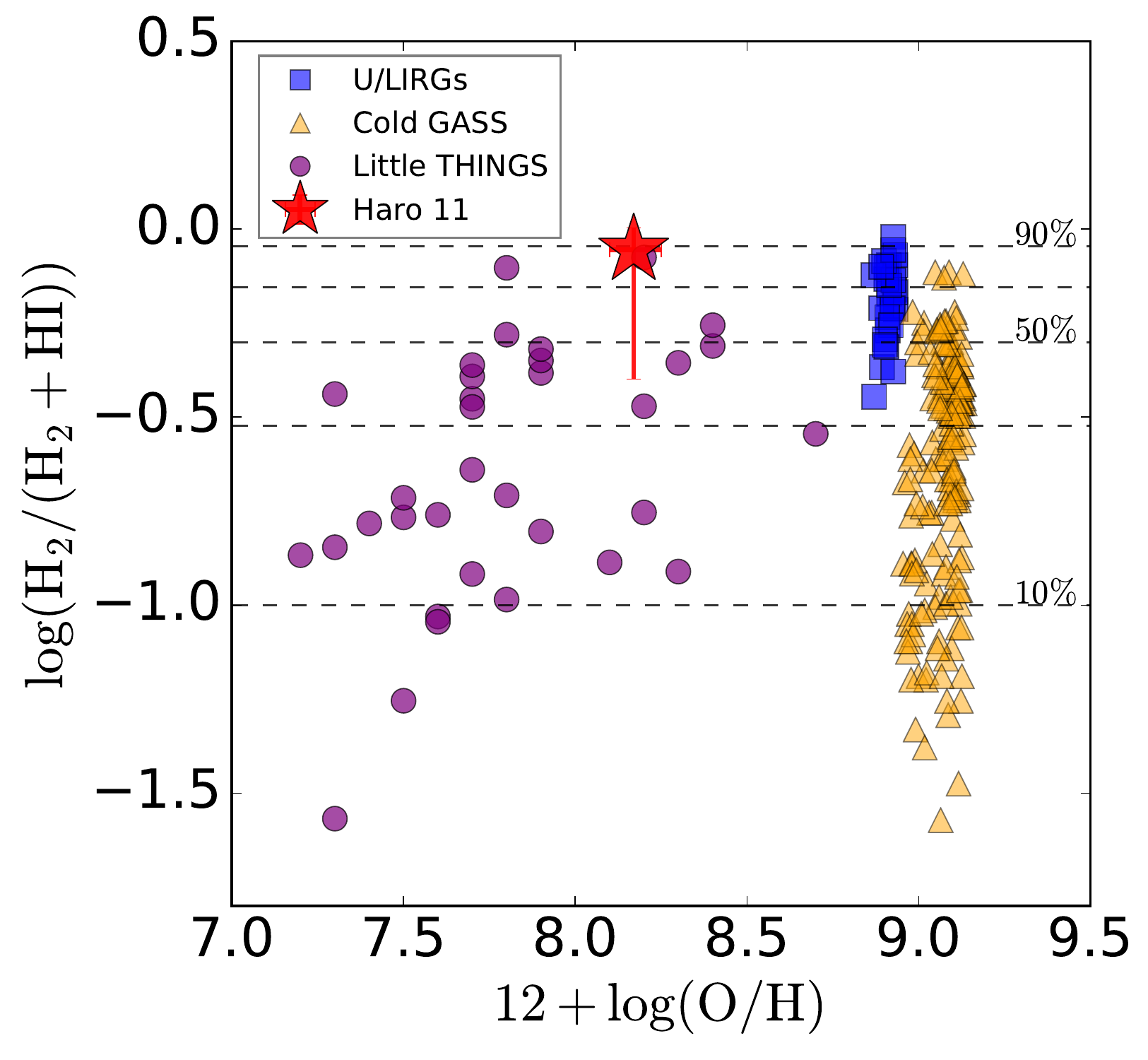}
   \includegraphics[width=0.45\textwidth]{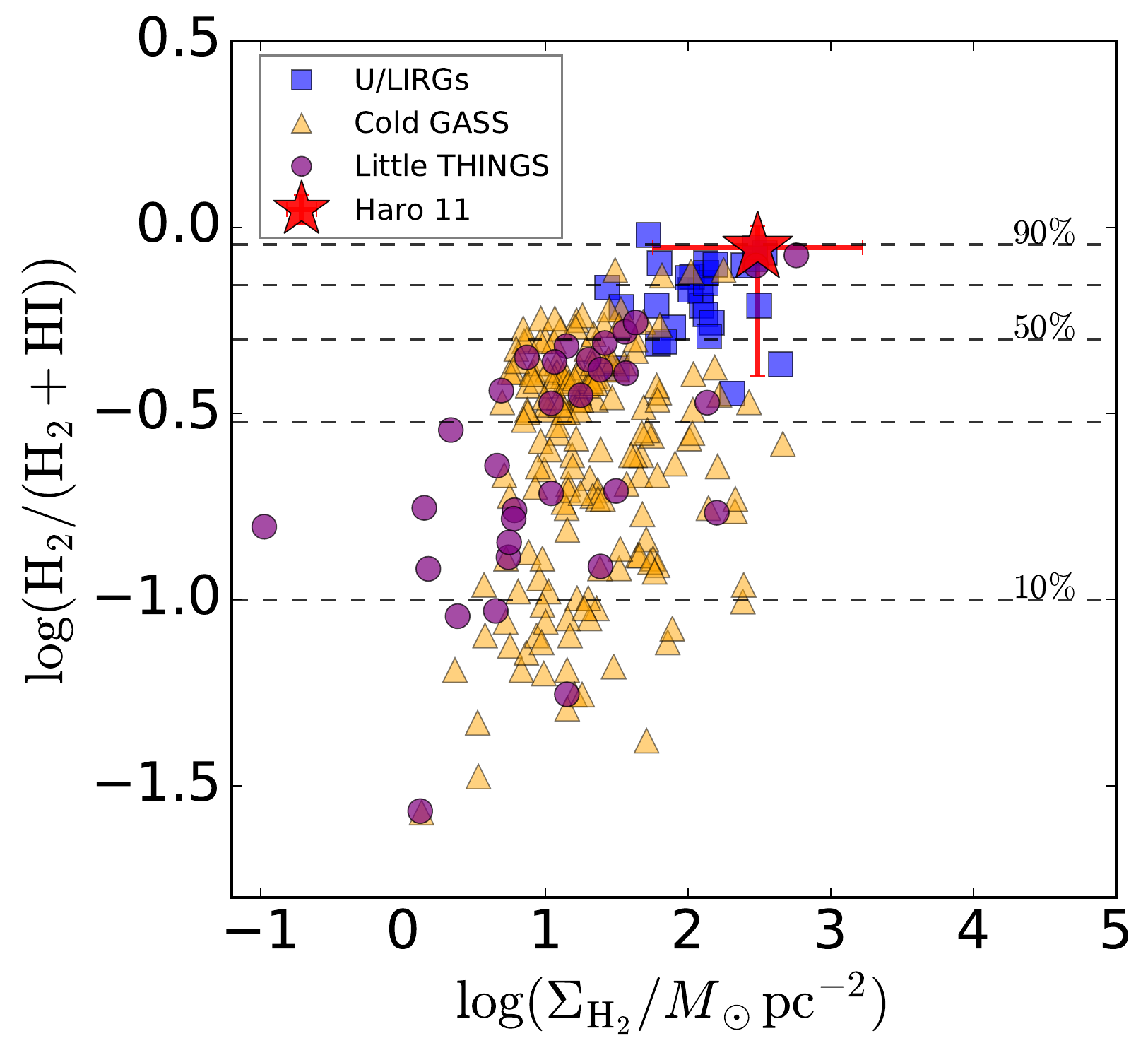}
   \caption{Molecular gas fraction ($M_{\rm H_2} / (M_{\rm H_2} + M_{\hi})$) as a function of the average gas phase metallicity ($left$ panel) and the average molecular gas surface density ($right$ panel) in Haro 11 (red star) and other nearby galaxies. The (U)LIRGS sample is marked as blue squares, which is collected from the GOALs survey \citep[e.g., ][]{armus2009,stierwalt2013,Shangguan2019a} and cross-matched with the ALFALFA $\hi$ survey \citep{Haynes2018a}, similar to the data in $right$ panel of Fig. \ref{fig:KS_law}. black triangles represent normal star forming galaxies with $M_* \geq 10^{10}\ \msun$ from Cold GASS survey \citep{Catinella2010,Catinella2012,Catinella2013}. The nearby dwarf irregular or compact galaxies collected from the Little THINGS survey \citep[e.g.,][]{Hunter2012,Hunter2019b} are shown as purple circles.}
   \label{fig:H2_HI}
   \end{figure*}

\subsection{Neutral gas}
\label{subsec:hi}

As the fuel of the molecular gas, neutral hydrogen gas is a prominent component of star-forming galaxies. To predict the evolution path of a merger system such as Haro 11, we need to consider the $\hi$-H$_2$ transition and the fraction of $\hi$ gas.  The spatial distribution and the kinematics of the $\hi$ gas are also essential to unravel the complex behaviors of molecular gas and star formation activities in Haro 11.  However, based on the 24 hours of observation with the VLA telescope (program 17B-287), \cite{2019AAS...23336810T} did not detect $\hi$ emission, confirming the surprising deficient neutral gas component within such a starburst galaxy \citep{Pardy2016}.  The transition from $\hi$ to H$_2$ is affected by the gas density, dust, metallicity, and radiation in the interstellar medium \citep[e.g., ][]{Sternberg2021}. This section will compare the fraction of $\hi$ and H$_2$ in Haro 11 with normal SFGs, massive (U)LIRGs, and other dwarf galaxies, then investigate the possible reasons for these differences.

In Fig. \ref{fig:H2_HI}, we show the molecular gas fraction ($M_{\rm H_2} / (M_{\rm H_2} + M_{\hi})$) as a function of the average gas-phase metallicity and the average molecular gas surface density in Haro 11 and other nearby galaxies. These nearby galaxies include the normal SFGs at $M_* \geq 10^{10}\ \msun$ in the Cold GASS survey \citep{Catinella2010,Catinella2012,Catinella2013}, massive merger (U)LIRGs in GOALS survey \citep[e.g., ][]{armus2009,stierwalt2013,Shangguan2019a} with the $\hi$ detection from ALFALFA survey \citep{Haynes2018a}, as well as the dwarf irregular and compact galaxies from the Little THINGS survey \citep[e.g.,][]{Hunter2012,Hunter2019b}. The molecular gas in the Cold GASS survey is determined from the CO(1-0) measurements from the IRAM 30-m telescope \citep{Catinella2010}. \cite{Shangguan2019a} performed the photometric spectral energy distribution fitting for (U)LIRGs to derive the total dust masses, then adopted the gas-to-dust ratios to estimate the gas masses. Their gas mass estimations are consistent with the molecular gas derived from CO measurements for a subsample of GOALS objects \citep{larson2016a,Shangguan2019a}. The molecular content of dwarf galaxies from Little THINGS survey is calculated from the product of the total SFR and the typical molecular consumption time \citep[$\sim$ 2 Gyr, ][]{bigiel2008}, following the method in \cite{Hunter2019b}. Meanwhile, we note that the gas consumption time of dwarf galaxies has a large uncertainty, ranging from $\sim$ 50 Myr to $\sim$ 10 Gyr \citep[e.g.,][]{hunt2015}. Using the consumption time of 50 Myr and 10 Gyr, the H$_2$ gas fraction of Little THINGS galaxies will be less than 10$\%$ and ranging from 10$\%$ to 90$\%$, respectively. However, the large uncertainty of the H$_2$ gas fraction can not change the trend that the gas fraction is dependent on the gas surface density, which we will mention in the next paragraph.

The metallicities of (U)LIRGs and normal SFGs are estimated from the stellar mass-metallicity relation (MZR) of \cite{Tremonti2004}, while are systemically $\sim$0.2 dex below the MZR in the merger (U)LIRGs \citep{herrera-camus2018}. Meanwhile, to compute the average molecular gas surface densities of these galaxies, we use the python packages \textit{Statmorph}\footnote{\url{https://statmorph.readthedocs.io/en/latest/overview.html}} \citep{rodriguez-gomez2019} and \textit{Photutils}\footnote{\url{https://photutils.readthedocs.io/en/latest/index.html}} to calculate the radius ($R_{\rm 80}$) containing $80\%$ flux in $V$ and $r$ band images. Within these sub-solar metallicity dwarf galaxies, we note the molecular gas fraction ($\sim 80\%$) of Haro 11 is rare, significantly larger than in normal SFGs, while is similar to the fraction ($\sim 50\% - 90\%$) of (U)LIRGs. At the similar metallicity, the molecular gas fraction in normal SFGs covers a large range ($\sim 5\% - 70\%$), probably indicating the average metallicity is not the major property to affect the molecular gas fraction and the $\hi$-H$_2$ transition in the ISM. However, if compared with the average H$_2$ surface densities, the H$_2$ fraction is larger with higher gas densities among different types of galaxies. This result also supports the scenario that the $\hi$-H$_2$ transition will be accelerated by the gravitational collapse in dense and optically thick regions \citep[e.g.,][]{sternberg2014, Sternberg2021}. Though the average gas surface density of Haro 11 is about $300\ \msun \ \rm pc^{-2}$, sightly higher than (U)LIRGs, the molecular gas is concentrated toward knots B and C regions, harboring high surface density molecular gas with $\Sigma_{\rm H_2} > 660\ \msun \ \rm pc^{-2}$. 

Based on the deficient $\hi$ gas, intense starburst activities (higher SFE), and the potential stellar/AGN feedback, we estimate the upper limit values of gas depletion time and the final stellar mass of Haro 11 at the evolutionary end-stage. The gas depletion time $(M_{\rm H_2} + M_{\hi}) / \rm SFR$ will be shorter than 0.17$^{+0.13}_{-0.12}$ Gyr. Assuming that all of the molecular and neutral gas will form stars, the final stellar mass $M_* + 1.36 (M_{\rm H_2} + M_{\hi})$ will be smaller than $8.5^{+5.2}_{-5.0} \times 10^9 \ \msun$, including the factor 1.36 to account for the helium. \cite{Oestlin2015} reported that Haro 11 shares similar morphology and kinematics of stars and ionized gas to those of the Antennae galaxy. After performing a high-resolution smoothed particle hydrodynamic simulation, \cite{Lahen2017} reproduced both the observed morphology and the off-nuclear starburst of the Antennae galaxy. \cite{Lahen2017} also found that the Antennae galaxy will evolve into a red quiescent galaxy after 2.5 Gyr of secular evolution in their simulation. Given the lower stellar mass and the shallow gravitational potential of Haro 11, the ionized and molecular gas will escape the host galaxy efficiently, which will reduce the molecular gas depletion time. We predict that Haro 11 can be the potential progenitor of these nearby less massive elliptical galaxies. In the $right$ panel of Fig. \ref{fig:KS_law}, we also point out the position of Haro 11 (marked as a white star) at the transition phase between SFGs and quiescent galaxies, where the stellar mass and SFR are assumed as their upper limit values. The sSFR at the transition phase is adopted as 10$^{-11}$ yr$^{-1}$ \citep[e.g.,][]{Renzini2015}, which is similar to these post-starburst galaxies from \cite{smercina2021} and green valley galaxies. In the future, we will perform a hydrodynamic particle simulation to reproduce the observed metallicity, star formation, gas fraction, as well as the kinematics of gas and stars in Haro 11, aiming to investigate the merger remnants at different evolutionary stages. 

%-----------------------------------------------------------------

\section{Summary}
\label{sec:sum}

In this work, using the archival ALMA (band 3, 8) and VLT/MUSE data, we study the spatially-resolved molecular gas and neutral carbon gas in the dwarf merger starburst galaxy Haro 11. We explore the star formation and the gas consumption activities in different regions. The main conclusions are summarized below.

1. Molecular gas is assembled around the central star-forming regions (knots B and C), while no significant detection of CO(1-0) is found within knots A and T. The molecular and ionized gas is approaching (receding) around knot C (B), while the stellar velocity is non-circular. Given the high gas velocity dispersion, these features indicate that the gas at knots B and C might be the complex/combined stage of collision of clouds and feedback from star formation.

2. High spatially-resolved SFEs and sSFRs are detected toward knot B. An offset between the local peaks of $\ha$ and CO emission is found at the bridge of B and C, and the gas therein will provide the fuel of the future star formation. Limited by the sensitivity of the observations, no significant CO is detected within knots A and T while indicating relatively higher sSFRs. This suggests the scenario that the molecular gas within A and T have probably been depleted by star formation or stellar feedback.

3. The peak location and distribution of [CI](1-0) strongly resembles the CO(1-0), indicating it could trace the same molecular gas as CO in such a dwarf merger starburst galaxy. The line ratios of [CI]/CO at the north region of knot C are about 0.5, much higher than ratios in Milky Way and normal spiral galaxies, probably caused by the dissociation of CO molecules by cosmic rays and FUV photons.

4. Haro 11 and the star-forming regions share similar SFEs and sSFRs as the high-$z$ starburst galaxies or the clumps in nearby (U)LIRGs. The turbulent pressure and viral parameter are significantly smaller than those in the massive merger galaxy NGC 4038/4039, (U)LIRGs, and post starburst galaxies, which will probably lead to the intense starburst activities in Haro 11. 

5. Given the high SFE, sSFR, small stellar mass, and low metallicity (1/3 $Z_{\odot}$), we argue that Haro 11 could be the analog of high-$z$ dwarf starburst galaxies and the potential progenitor of the nearby less massive elliptical galaxies. Considering the deficient $\hi$ gas and the potential stellar/AGN feedback therein, we predict that star formation in Haro 11 will become quenched at $M_* \leq 8.5 \times 10^9 \ \msun$ after a time scale less than 0.2 Gyr.

\begin{acknowledgements}
  We thank the referee Dr. Katie Jameson for thoughtful comments and insightful suggestions that improve our paper greatly.
  We thank Drs. Guilin Liu, Xu Kong and Xianzhong Zheng for fruitful discussions and advice.
  This work is supported by the National Key Research and Development Program of China (No. 2017YFA0402703), and by the National Natural Science Foundation of China (No. 11733002, 12121003, 12192220, and 12192222). We also acknowledge the science research grants from the China Manned Space Project with NO. CMS-CSST-2021-A05.
  Y.L.G acknowledge the grant from the National Natural Science Foundation of China (No. 12103023). M.H. is fellow of the Kunt $\&$ Alice Wallenberg foundation.
  This paper makes use of the following ALMA data: ADS/JAO.ALMA$\#$ 2013.1.00350.S and 2017.1.01457.S. ALMA is a partnership of ESO (representing its member states), NSF (USA), and NINS (Japan), together with NRC (Canada) and NSC and ASIAA (Taiwan), in cooperation with the Republic of Chile. The Joint ALMA Observatory is operated by ESO, AUI/NRAO, and NAOJ. The National Radio Astronomy Observatory is a facility of the National Science Foundation operated under cooperative agreement by Associated Universities, Inc.
  This research has made use of the NASA/IPAC Extragalactic Database (NED), which is funded by the National Aeronautics and Space Administration and operated by the California Institute of Technology.

\end{acknowledgements}

\bibliographystyle{aa}
\bibliography{Haro11}

\end{document}